%% file: PH_anatomy.tex
\documentclass[a4paper,11pt]{article}

\usepackage[normalem]{ulem}
\usepackage{xcolor}
\usepackage[english]{babel}
\usepackage[T1]{fontenc}
\usepackage[utf8]{inputenc}
\usepackage{authblk}
\usepackage{mathtools}
\usepackage{epsfig}
\usepackage{slashed}
\usepackage{amsmath,amssymb}
\usepackage{mathrsfs}
\usepackage{amsfonts}
\usepackage{enumitem}
\usepackage{graphicx,color,xcolor}
\usepackage{cite}
\usepackage{float}
\usepackage{subcaption}
\usepackage{soul}
\usepackage{hyperref}
\usepackage{wrapfig}
\usepackage{booktabs}
\usepackage{array}
\usepackage{microtype}
\usepackage{bm}
\usepackage{cleveref}
\usepackage[ruled,vlined]{algorithm2e}
\usepackage[left=2.5cm,right=2.5cm,top=2.5cm,bottom=2.5cm]{geometry}
\numberwithin{equation}{section}

\definecolor{refcol}{rgb}{0.9,0.1,0.1}
\hypersetup{colorlinks=true,linkcolor=blue,citecolor=refcol,urlcolor=cyan,linktocpage}
\newcommand{\pythia}{\textnormal{\textsc{Pythia}}}
\newcommand{\herwig}{\textnormal{\textsc{Herwig}}}
\newcommand{\mg}{\textnormal{\textsc{MadGraph5\_aMC@NLO}}}

\newcommand{\auc}{\mathrm{AUC}}
\newcommand{\ess}{N_{\mathrm{eff}}}
\newcommand{\dd}{\mathrm{d}}

\newcommand{\Ccommon}{\mathcal{C}_{\mathrm{common}}}

\input{numbers.tex}

\begin{document}

\begin{titlepage}
\thispagestyle{empty}

\title{{\huge\bf Functional anatomy of Pythia–Herwig differences with Kolmogorov–Arnold networks}}

\vfill

\author{
{\bf Arghya Chattopadhyay}\thanks{{\tt arghya.chattopadhyay@upr.edu}}
\smallskip\hfill\\
\small{
{\it Department of Physics, University of Puerto Rico at Mayag\"uez}\\
{\it Mayag\"uez, Puerto Rico 00681, USA}
\vspace{1cm}
}
}
\bigskip\bigskip\bigskip\bigskip
\vfill
\date{
\begin{quote}
\centerline{{\bf Abstract}}
{\small
Differences between high-energy event generators can arise at several stages of the collision simulation, from the hard scattering through parton showering and hadronization to the final event. These differences are usually summarized using observable distributions or global classifier scores. While these quantify the disagreement, they do not reveal which observable-level structures carry it or whether those structures persist through different stages of event generation. In this work, we formulate this problem as a staged functional analysis of generator-model differences. Following the same hard dijet events through \textnormal{\textsc{Pythia}} and \textnormal{\textsc{Herwig}} at shower-only, hadronized, and full-generator levels, we use an additive Kolmogorov-Arnold network (KAN) representation of the classifier-derived log density ratio to decompose the learned discrepancy into explicit one-dimensional observable responses that can be isolated, recomposed, and transported between generator stages. Within the same eight-observable jet representation, the \textnormal{\textsc{Pythia}}-\textnormal{\textsc{Herwig}} difference is driven mainly by multiplicity at shower level, shifts toward jet mass and shape after hadronization, and develops a mixed shape-multiplicity driven structure in the full-generator configuration. Transporting the individual shower-level functional components downstream shows that shower-level multiplicity information can retain its reweighting power, whereas the corresponding shape responses need not do so even though shape becomes important again at later stages. The jet-mass factors, meanwhile, are limited by poor statistical support. This KAN-based framework therefore provides a functional anatomy of generator-model dependence, exposing both persistent structures and support failures that are hidden inside a single global classifier-derived reweighting function.
}
\end{quote}}
\end{titlepage}

\thispagestyle{empty}
\maketitle
\vfill
\eject

\tableofcontents

\section{Introduction}
\label{sec:intro}
To validate any experimental observations, one needs to perform a theory-driven simulation first to make sense of the experimental data, which we only get in the form of detector-level signals. Event generators are therefore a crucial ingredient of collider phenomenology, combining our best theoretical understanding of fundamental interactions with phenomenological models for those aspects of the collision that cannot yet be inferred entirely from first principles. A complete event generation proceeds in stages, beginning with the short distance partonic scattering computed from field theory amplitudes, followed by a parton shower that organizes the dominant soft and collinear radiation, and by hadronization, which converts the colored partonic state into color-singlet hadrons. Finally, multiparton interactions (MPI), beam remnants, and related modeling build the underlying event. General-purpose Monte Carlo generators combine these ingredients into complete simulated events, but there is no unique first-principles prescription for all stages. Differences between independent generators, such as \pythia{} and \herwig{}, can therefore be used to probe how alternative modeling choices manifest themselves in the resulting event description.

The path from an underlying hard interaction dictated by some well-motivated quantum field theory lagrangian to a fully simulated high-energy collision is not unique, mostly because of the interplay between perturbative and non-perturbative physics. Different event generators implement the perturbative and non-perturbative stages using different, but physically well-motivated, prescriptions. In particular, \pythia{} and \herwig{} differ substantially in their treatment of parton showering and hadronization. \pythia{} conventionally employs a transverse-momentum ordered shower together with the Lund string model \cite{Bierlich:2022pfr,Andersson:1983ia}, while \herwig{} is traditionally built around angular-ordered showering and cluster hadronization \cite{Bewick:2021nhc,Webber:1983if}. These differences reflect alternative ways of modeling physics in regions where perturbative calculations alone do not suffice. It is therefore natural, and phenomenologically important, to compare different generators and ask which physical features are responsible for their different predictions.

Recently, machine learning (ML) based density-ratio estimation has provided a way to formulate this comparison directly as an event-by-event reweighting problem \cite{Cranmer:2015bka,Andreassen:2019nnm}. In particular, \cite{Furuichi:2025generator} demonstrated neural re-weighting between \pythia{} and \herwig{} using high-level observables, while also highlighting practical difficulties associated with large event weights and incomplete observable representations. Such a comparison, however, becomes particularly subtle if the learned difference is interpreted as a theory uncertainty. Ghosh and Nachman showed that reducing the apparent difference between two theory predictions through decorrelation does not necessarily imply a corresponding reduction of the underlying theory uncertainty \cite{Ghosh:2021hrh}. This motivates going beyond a single classifier score or global reweighting function and asking instead which physical observables carry the \pythia{}-\herwig{} difference, how these contributions evolve through showering and hadronization, and whether the structures identified at an earlier stage remain meaningful in the final event.

The present work addresses this missing layer of information. We construct a controlled staged data set in which the same hard $pp\to jj$ events are evolved with \pythia{} and \herwig{} through three controlled generator configurations: shower only, shower plus hadronization, and the full generator including MPI and the underlying event. We then restrict the analysis to a common eight-observable dijet representation that is defined for both generators at all three stages. To get a properly justified comparison, the hard-event identity, event population, jet algorithm, and observable definitions are therefore held fixed while the generator evolution changes.

Kolmogorov-Arnold networks (KANs) provide a particularly useful language for this problem. Previous HEP studies have investigated the application of KANs to classification and interpretability \cite{Abasov:2024kan}. In particular, Erdmann et al. showed that a single-layer KAN can learn functions resembling one-dimensional log-likelihood-ratio contributions \cite{Erdmann:2025kan}. KANs replace the scalar edge weights of a conventional multilayer perceptron (MLP) by trainable one-dimensional functions, usually represented numerically by splines \cite{Liu:2024kan}. An MLP can certainly be trained as a generator classifier and therefore used for density-ratio reweighting. The basic KAN and MLP difference lies in their structure, contrary to a MLP, which is parameterized by dense collections of weights and biases acting through fixed activation functions, KAN live in a functional space of one dimensional splines. Rather than using KAN for generic accuracy advantage over MLPs, the feature we exploit here is precisely this structural difference. Since for a deliberately additive KAN the learned logit can be written as
\begin{equation}\label{eq:add_KAN}
  \ell_{\mathrm{add}}(x)=b+\sum_i f_i(x_i),
\end{equation}
therefore every observable-dependent function $f_i$ become an explicit, exportable object and the corresponding weight factorizes exactly within the fitted model. This allows us to study the combination of native observable-level factorization with a stage-by-stage \pythia{}-\herwig{} comparison and direct transport of individual learned factors through the event-generation chain.  

Equation \eqref{eq:add_KAN} gives a particularly transparent way of using the input observables $x_i$ themselves to understand how the generator difference is built. Since each input observable is associated with an explicit response function $f_i(x_i)$, one can then \emph{open} the KAN and inspect these individual functions rather than treating the classifier as a single black-box output. This simple observation then allows us to ask directly which jet observables carry the \pythia{}-\herwig{} difference, over which kinematic regions they do so, and how the different observable contributions combine.

This motivates the two complementary analyses performed in this work. \textbf{Analysis I} studies the structure of the \pythia{}-\herwig{} difference separately at the shower-only, hadronized, and full-generator stages. We first compare the additive KAN with a deeper KAN to determine how much of the separation can already be described through the individual observable responses $f_i(x_i)$. We then use the explicit additive decomposition, together with subset recomposition and Shapley allocation\footnote{For more details, see \cref{app:shapley}.}, to quantify which observables are responsible for the generator separation at each stage. \textbf{Analysis II} asks whether a difference identified at the shower level retains any downstream reweighting power. In this case, an individual factor $f_i(x_i)$ learned from the shower-level events is used directly as an event-by-event reweighting factor when studying the corresponding hard events after hadronization and in the full-generator configuration. This is where the KAN representation is particularly useful, the contribution associated with a specific physical observable is already an explicit part of the learned function and can therefore be isolated and followed through the subsequent stages of event generation.

The three-stage construction is chosen to localize how the observable structure of the \pythia{}-\herwig{} difference changes through successive stages of event generation. Each stage corresponds to a controlled generator configuration with a well-defined role in the event-generation chain, while the hard events, event selection, and eight observable representation are kept fixed throughout. This allows us to study how the observable content of the learned generator difference reorganizes as hadronization and the full event structure are introduced, without simultaneously changing the underlying hard process or the observable basis. In this controlled setting, we can ask whether the difference can be resolved into explicit observable level functions and more importantly, whether any of these structures remain meaningful at later stages of event generation. 

The paper is organized as follows. \Cref{sec:data} defines the hard process, generator stages, jet reconstruction, and common data set. \Cref{sec:kanmethod} develops the KAN and density-ratio formalism, including training, cross-fitting, exact functional recomposition, and Shapley allocation. \Cref{sec:analysisI} presents the stage-dependent anatomy, while \cref{sec:analysisII} performs source-to-target factor transport. \Cref{sec:conclusion} summarizes the conclusions, limitations, and possible directions with this KAN-based-framework. Technical implementation, statistical definitions, exact Shapley construction, and weight-support diagnostics are collected in the appendices. The analysis code and complete run instructions are collected in the github repository \cite{Anatomy_pythia_herwig}.

\section{Hard process and generator stages}
\label{sec:data}

This section describes the data-generation procedure used throughout the two analyses, together with the main choices, cuts, and stage definitions entering the construction of the sample. The main requirement is that the same underlying hard scattering should be followed through the different generator stages while keeping both the event identity and the observable basis fixed. This is important physically because otherwise a change observed between the shower-only, hadronized, and full-generator samples could arise either from the newly introduced physics or simply from comparing different hard events or different representations. By controlling these ingredients, the staged comparison can be interpreted more directly as showing how the \pythia{}-\herwig{} difference evolves as additional layers of event generation are introduced.

\subsection{Common hard events and generator configurations}
\label{sec:hardprocess}

We generate leading-order $pp\to jj$ events at $\sqrt{s}=13~\mathrm{TeV}$ with \mg{} \cite{Alwall:2014hca}.  The outgoing matrix-element partons satisfy
\begin{equation}
 p_T>20~\mathrm{GeV},\qquad |\eta|<5,\qquad \Delta R_{jj}>0.4.
\end{equation}
Here, $p_T$ denotes the momentum transverse to the beam direction, $\eta=-\ln\tan(\theta/2)$ is the pseudorapidity, with $\theta$ measured from the beam axis, and $\Delta R_{jj}=\sqrt{(\Delta\eta)^2+(\Delta\phi)^2}$ measures the angular separation between the two outgoing partons in the pseudorapidity-azimuth plane. The same Les Houches event (LHE) records \cite{Alwall:2006yp} are subsequently supplied to \pythia{} 8.312 \cite{Bierlich:2022pfr} and \herwig{} 7.3.0 \cite{Bewick:2023tfi}. Consequently, the matrix-element event remain common to both the generators, comparison begins only after the hard scattering is fixed. For each generator we construct three configurational stages:
\begin{description}[leftmargin=1.55cm,style=nextline]
\item[Stage A:] initial and final state parton showering is enabled, while hadronization and MPI are disabled. The resulting events therefore contain the showered partonic final state only;
\item[Stage B:] hadronization is added on top of each generator's own parton
shower, while MPI and underlying-event activity remain disabled;
\item[Stage C:] the full generator configuration is used, including showering,
hadronization, MPI, and underlying-event modeling;
\end{description}
where underlying-event modelling points toward the additional, typically softer, hadronic activity accompanying the primary hard scattering, arising mainly from multiparton interactions, beam remnants, and their subsequent hadronization. These differences make the comparison physically meaningful, since the two generators realize the evolution from the same hard scattering to the final state through different, well-established modeling choices. The \pythia{} branch uses a $p_T$-ordered parton shower together with Lund-string hadronization, while the
\herwig{} branch uses an angular-ordered shower followed by cluster hadronization. Throughout all three stages, \pythia{} is run with the Monash 2013 tune \cite{Skands:2014pea}, while \herwig{} 7.3 is used with its default parameter configuration \cite{Bewick:2023tfi}.

The staged setup allows one to identify at which point in the event-generation chain the \pythia{}-\herwig{} difference changes, but it should not be interpreted as isolating a single microscopic ingredient. The transition from Stage A to Stage B, for example, probes how the \pythia{}-\herwig{} difference changes when hadronization is introduced. It should not, however, be interpreted as a controlled comparison of Lund-string and cluster hadronization acting on an
identical shower, since the two generators already produce different showered partonic states at Stage A.

The three stages are generated as separate controlled configurations starting from the same LHE hard event. The common \texttt{hard\_event\_id} therefore links the underlying hard scattering across stages A, B, and C, but the analysis does not require a common stochastic shower history or a uniquely matched event trajectory between the stages. The staged comparison is defined
at the level of the common hard-event identity rather than by following an individual reconstructed jet through the generator chain.

\subsection{Jet reconstruction at each stage}
\label{sec:jets}

Jets are reconstructed independently for both generators at each of the three stages, resulting in six reconstructed jet representations.  The objects passed to FastJet depend on the generator stage. At Stage A, FastJet clusters the final partons produced by the parton shower. At Stages B and C, FastJet instead clusters the stable visible final-state particles produced by the generator, with neutrinos removed. In every case we use FastJet \cite{Cacciari:2011ma} with the anti-$k_T$ algorithm \cite{Cacciari:2008gp}, radius $R=0.4$, and E-scheme recombination\footnote{In the E-scheme, the four-momenta of clustered constituents are simply added, so the jet four-vector is the sum of the four-vectors assigned to it by the clustering sequence.}.

At this point we should emphasize an important point about the staged comparison in the present study. The jets under consideration themselves are not tracked or matched from A to B to C. FastJet is rerun separately on the final state at each stage, and the leading and subleading jets are defined afresh by reconstructed $p_T$.  What is tracked is the underlying \emph{hard event}, it does not mean following a uniquely matched jet through hadronization.

The event-level dijet selection is
\begin{equation}
 p_{T,j_1}\ge 25~\mathrm{GeV},\qquad
 p_{T,j_2}\ge 20~\mathrm{GeV},\qquad
 |y_j|\le 4.4,
 \label{eq:jet_selection}
\end{equation}
with $y$ being rapidity.

\subsection{The eight-observable representation}
\label{sec:features}

The common representation of choice for this analysis is the set of parameters
\begin{equation}
 x=\left(n_1,m_1,p_{T,1}^{D},g_1,n_2,p_{T,2}^{D},m_2,g_2\right),
 \label{eq:features}
\end{equation}
where the subscripts $1$ and $2$ denote the leading and subleading reconstructed jets.  The constituent multiplicity $n_j$ is the number of FastJet input objects assigned to the particular jet corresponding the particular stage in consideration. It is therefore a well-defined generator diagnostic at every stage, but its microscopic meaning changes when the object definition changes from partons to hadrons.

The jet mass $m$ is computed from the summed four-vector of all jet constituents. Even if the individual constituents are massless, the jet can acquire a nonzero invariant mass because their momenta are not exactly collinear. $p_T^D$ quantifies whether the transverse momentum is shared among many constituents or concentrated in a few, and the girth $g$ is a $p_T$-weighted radial spread of the jet \cite{Gallicchio:2011xq}, both of which can be mathematically written down as
\begin{equation}
 p_T^D=\frac{\sqrt{\sum_{i\in j}p_{T,i}^2}}{\sum_{i\in j}p_{T,i}},
 \qquad
 g=\frac{\sum_{i\in j}p_{T,i}\,\Delta R_{iJ}}{\sum_{i\in j}p_{T,i}},
 \label{eq:ptd_girth}
\end{equation}
where $\Delta R_{iJ}$ is the distance between constituent $i$ and the jet axis in rapidity-azimuth space. Constituent multiplicity and $ p_T^D$ are not infrared nor collinear safe observables, multiplicity is sensitive to both soft emissions and collinear splittings, while  $p_T^D$ is particularly sensitive to collinear splittings \cite{Larkoski:2014pca}. This is not problematic for
the present analysis, because here they are used as generator diagnostics rather than as observables for a standalone fixed-order perturbative prediction.

Apart from computational simplicity, the restriction to eight observables is motivated first by the requirement of \emph{common definedness}. Since we want to follow the same hard event through all generator stages, every observable used in the analysis should remain well defined in all six reconstructed representations. This becomes non-trivial at Stage A, where some jets can contain only a very small number of shower partons. More elaborate substructure observables may require several resolved constituents or involve ratios whose denominator can vanish in such sparse jets. Requiring those observables would therefore remove different events from the \pythia{} and \herwig{} samples and would change the common hard-event cohort itself, as discussed in the next subsection. The retained variables span constituent counting, invariant mass, momentum sharing, and radial structure while remaining defined for the common cohort. For example, even a single-constituent jet has a well-defined multiplicity, $p_T^D=1$, $g=0$, and a well-defined jet mass, which is zero for a single massless constituent.

There is also a practical advantage to keeping this representation between a small number of variables. As discussed in the next section, the additive KAN allows the contribution of each observable to be isolated and different combinations of these contributions to be reconstructed directly from the trained model. With eight observables there are $2^8=256$ possible subsets, including the empty and full sets, so every possible combination can be studied exactly without retraining the network. This exhaustive decomposition will later be used to quantify how the generator separation is distributed among the observables. The choice of eight observables is therefore especially convenient for the analysis, although the primary motivation remains the requirement that the same representation be well defined throughout the generator chain, allowing for more general ocnsiderations in the future.

\subsection{Building the common hard-event cohort}
\label{sec:cohort}

The purpose of the three-stage construction is to compare how \pythia{} and \herwig{} evolve the same underlying hard scatterings. For this reason, we construct a common hard-event cohort, denoted by $\Ccommon$, which contains only those LHE events that can be followed consistently through every generator and every stage. A hard event is included in $\Ccommon$ only if
\begin{enumerate}[label=(\roman*)]
	\item its identity is recovered in all six generator-stage outputs;
	\item it passes the selection in \cref{eq:jet_selection} in all six
	reconstructed representations; and
	\item all eight observables in \cref{eq:features} are finite and physically
	defined in all six representations.
\end{enumerate}
These requirements are designed to serve different purposes. The first establishes the event-by-event correspondence needed for the staged comparison. A \pythia{} event and a \herwig{} event carrying the same hard-event identifier originate from the same LHE hard scattering, so differences observed later arise from their subsequent generator evolution rather than from comparing different hard-process events.

The second requirement fixes the event population on which the comparison is made. Showering, hadronization, and MPI can modify the reconstructed jets, so an event that satisfies the dijet selection in one representation need not do so in another. If each sample were selected independently, the classifier could therefore learn not only differences in the jet observables but also differences caused by the six samples containing different hard events. Requiring the same dijet selection in all six representations removes this ambiguity by restricting the analysis to their common acceptance protocol. Consequently, generator-dependent acceptance differences outside this intersection are not part of the density ratio studied below.

The third requirement similarly ensures that every accepted hard event can be described by exactly the same set of eight-dimensional observable vector throughout the analysis. No observable is imputed and no event is retained in one generator while being removed from another because one of its observables is undefined. This requirement is particularly relevant at Stage A, where sparse partonic jets can make more elaborate jet-substructure observables ill-defined, as discussed in the last subsection.

The resulting $\Ccommon$ therefore represents a fixed set of hard scatterings for which both generators produce a valid selected dijet event at all three stages and for which the same observable representation can always be constructed. This common-cohort requirement is central to this analysis, since it allows changes between stages to be studied without simultaneously changing the underlying hard-event population.

In the production sample used for the results presented in this paper, we start from $\NCandidate$ candidate LHE events and obtain $\NCommon$ events satisfying all three requirements, corresponding to a common-cohort efficiency of $\CommonEff$. The hard-event identifiers are then split deterministically into $\NDev$ development events and $\NTest$ held-out final-test events. All six representations of a given hard event are always kept in the same partition. Within the development sample they are also assigned to the same cross-validation fold. This prevents different representations of the same underlying hard scattering from entering the training and validation samples simultaneously. The technical implementation of the event identity and the corresponding consistency checks are described in \cref{app:technical}.

The common-cohort construction also determines precisely which probability
distributions are compared. Let $p_{P,s}(x\mid\Ccommon)$ and $p_{H,s}(x\mid\Ccommon)$ denote the normalized probability densities of the eight-dimensional observable vector $x$ obtained with \pythia{} and \herwig{}, respectively, at generator stage $s\in\{A,B,C\}$, after restricting the event population to the common cohort $\Ccommon$. The density-ratio target at each stage is therefore
\begin{equation}
	r_s(x)
	\equiv
	\frac{p_{H,s}(x\mid\Ccommon)}
	{p_{P,s}(x\mid\Ccommon)} .
	\label{eq:conditional_target}
\end{equation}
This ratio measures how differently \herwig{} and \pythia{} populate the same eight-dimensional observable space when evaluated on the same selected population of underlying hard events.  With equal \pythia{} and \herwig{} \emph{class priors}, the connection between this quantity and the optimal classifier output is derived in the next section.

Since both $p_{P,s}$ and $p_{H,s}$ are normalized within $\Ccommon$, $r_s(x)$ describes differences in the \emph{shape of the observable distribution} within the common cohort. It is therefore not an inclusive \herwig{}/\pythia{} cross-section ratio. Differences associated with the overall event yield, generator failures, or events that fall outside the common acceptance are not included in the density ratio defined in \cref{eq:conditional_target}, making this a well-defined physics probe.

\section{Theoretical framework: KAN and functional reweighting}
\label{sec:kanmethod}

\subsection{From the Kolmogorov-Arnold theorem to a trainable network}
\label{sec:katheorem}

The Kolmogorov-Arnold representation theorem states that a continuous function of several variables on a compact domain can be represented through finite compositions of continuous one-dimensional functions and addition \cite{Kolmogorov:1957,Arnold:1957}. For a function of $d$ variables, the representation can be written schematically as
\begin{equation}
	F(x_1,\ldots,x_d)
	=
	\sum_{q=1}^{2d+1}
	\Phi_q\!\left(
	\sum_{p=1}^{d}\phi_{qp}(x_p)
	\right).
	\label{eq:KA_theorem}
\end{equation}
The importance of the theorem in the present context is that a genuinely multivariate function can, in principle, be constructed from functions of one variable. The theorem itself is an existence statement and does not prescribe how such functions should be learned from finite data.

Kolmogorov-Arnold networks (KANs) turn this basic idea into a trainable neural network architecture \cite{Liu:2024kan}. In a conventional multilayer perceptron (MLP), information is propagated using scalar weights followed by fixed nonlinear activation functions. A KAN instead assigns a trainable one-dimensional function to each edge of the network. A KAN layer can be
written as
\begin{equation}
	z^{(\ell+1)}_j
	=
	\sum_i
	\phi^{(\ell)}_{ji}\!\left(z^{(\ell)}_i\right),
	\label{eq:KAN_layer}
\end{equation}
where the functions $\phi^{(\ell)}_{ji}$ are learned during training.  In the pyKAN implementation in \cite{pykanRepo} used in this work, these functions contain trainable B-spline components together with a smooth base function. Thus, the essential difference from an MLP for our purpose is not simply the number of trainable parameters but the object that is being learned itself. For a more detailed comparison of KAN and conventional neural-network architectures the reader can refer to the work \cite{Liu:2024kan}.

For later analysis, the learned one-dimensional responses are exported as dense numerical tables over the range populated by the training data. These exports reproduce the corresponding trained KAN response within a verified numerical tolerance and allow the individual functions to be evaluated, removed, and recombined without retraining the network. One can therefore regard them as explicit numerical representations of the learned functions rather than as a truely elementary analytic formula.

For the central analysis we deliberately choose the simplest KAN architecture that exposes the dependence on the eight observables directly, which is a width $[8,1]$ network. Since each of the eight inputs is connected directly to a single output, its raw output is an additive function of the eight input observables. We also train a deeper width $[8,4,1]$ KAN. The hidden layer allows information from different observables to be combined before the final output and therefore provides a diagnostic of information that cannot be captured by the deliberately additive model. The deeper network is used only as a diagnostic of additional non-additive separation within this representation. 

One should note that, we do not use an MLP as a performance benchmark in this work. A conventional MLP can certainly be used for generator classification and density-ratio estimation. What matters here is not competitive classification performance, but that the
$[8,1]$ KAN makes the learned generator difference directly accessible at the level of individual observables.

\subsection{From a generator classifier to a density ratio}
\label{sec:ratio}

We now establish the connection between generator classification and reweighting.  At a fixed generator stage $s$, let $y=0$ denote an event generated with \pythia{} and $y=1$ an event generated with \herwig{}. Before looking at the observable vector $x$, the classification sample contains some fraction of events from each of these two classes. We denote these fractions or  \emph{class priors} by
\begin{equation}
	\pi_P=P(y=0),
	\qquad
	\pi_H=P(y=1),
	\qquad
	\pi_P+\pi_H=1.
	\label{eq:class_priors}
\end{equation}
In the present analysis each hard event contributes one \pythia{} and one \herwig{} representation. We therefore use equal class priors,
\begin{equation}
	\pi_P=\pi_H=\frac{1}{2}.
	\label{eq:equal_priors}
\end{equation}
As defined in the previous section, $p_{P,s}(x\mid\Ccommon)$ and $p_{H,s}(x\mid\Ccommon)$ are the normalized probability densities of the eight-dimensional observable vector for \pythia{} and \herwig{} at stage $s$, restricted to the same set $\Ccommon$. To keep the derivation compact, we write these densities from now onwards as $p_P(x)$ and $p_H(x)$. Suppose that a binary classifier assigns to an event the probability
\begin{equation}
	d(x)=P(y=1\mid x),
\end{equation}
that it belongs to the \herwig{} class. The population binary cross-entropy loss can be written as
\begin{equation}
	\mathcal{L}[d]
	=
	-\int \dd x\,
	\left[
	\pi_H p_H(x)\log d(x)
	+
	\pi_P p_P(x)\log\!\left(1-d(x)\right)
	\right].
	\label{eq:bce_population}
\end{equation}
Since the value of $d(x)$ at one point does not affect the integrand at another, the minimum can be found point by point. At fixed $x$,
\begin{equation}
	\frac{\partial\mathcal{L}_x}{\partial d}
	=
	-\frac{\pi_H p_H(x)}{d}
	+
	\frac{\pi_P p_P(x)}{1-d}
	=0,
\end{equation}
which gives the Bayes-optimal classifier
\begin{equation}
	d^*(x)
	=
	\frac{\pi_H p_H(x)}
	{\pi_H p_H(x)+\pi_P p_P(x)}.
	\label{eq:bayes_classifier_general}
\end{equation}
The odds associated with this classifier are therefore
\begin{equation}
	\frac{d^*(x)}{1-d^*(x)}
	=
	\frac{\pi_H}{\pi_P}
	\frac{p_H(x)}{p_P(x)}.
	\label{eq:bayes_odds_general}
\end{equation}
For the equal class priors in \cref{eq:equal_priors}, the prior factor cancels. The optimal classifier then becomes
\begin{equation}
	d^*(x)
	=
	\frac{p_H(x)}
	{p_P(x)+p_H(x)},
	\label{eq:bayes_classifier}
\end{equation}
and its logit is
\begin{equation}
	\ell^*(x)
	\equiv
	\log\frac{d^*(x)}{1-d^*(x)}
	=
	\log\frac{p_H(x)}{p_P(x)}
	=
	\log r_s(x),
	\label{eq:bayes_ratio}
\end{equation}
where $r_s(x)$ is the conditional density ratio defined in \cref{eq:conditional_target}. Thus, for equally represented classes, the Bayes-optimal classifier logit is exactly the logarithm of the
\herwig{}-to-\pythia{} density ratio on the common cohort \cite{Cranmer:2015bka}.

In the present implementation, the network itself returns the raw logit $\ell_s(x)$. The binary cross-entropy loss is evaluated directly from this logit, which is mathematically equivalent to applying $d(x)=1/(1+e^{-\ell_s(x)})$ before evaluating the cross entropy. Keeping the raw logit is particularly convenient here because it is the quantity that approximates the logarithm of the density ratio. A finite network trained on a finite data sample will not in general coincide with the Bayes-optimal function. Therefore one should rather interpret the trained $\ell_s(x)$ as a \emph{classifier-derived approximation} to the conditional log density ratio. Its ability to perform reweighting is tested explicitly in the subsequent analyses rather than assuming that \cref{eq:bayes_ratio} is an exact identity for the fitted network.

\subsection{Additive KAN and factorization of the event weight}
\label{sec:additive}

For the central width-$[8,1]$ KAN, the raw classifier logit has the additive form
\begin{equation}
	\ell_{s,\mathrm{add}}(x)
	=
	b_s+\sum_{i=1}^{8}f_{s,i}(x_i),
	\label{eq:add_kan}
\end{equation}
where $x_i$ denotes one of the eight jet observables and $f_{s,i}(x_i)$ is its learned one-dimensional response at stage $s$. This equation is the main reason for using the additive KAN in the present study. Instead of obtaining only one global classifier score, \eqref{eq:add_kan} allows one to inspect the individual function associated with each physical observable.

The separation between the constant term and the individual functions is not unique. For some arbitrary constants $c_i$, the transformation
\begin{equation}
	f_{s,i}(x_i)\rightarrow f_{s,i}(x_i)+c_i,
	\qquad
	b_s\rightarrow b_s-\sum_i c_i
\end{equation}
leaves the total logit unchanged. We fix this \emph{gauge-like} freedom\footnote{In the sense that different choices of the individual offsets represent the same total fitted function.} by centering each exported component on the reference development sample used for the exported fit,
\begin{equation}
	\mathbb{E}_{\mathrm{ref}}
	\!\left[f_{s,i}(X_i)\right]=0,
	\label{eq:center}
\end{equation}
and absorbing the corresponding offsets into $b_s$. This convention changes neither the classifier score nor the complete event weight. It simply gives a common reference level from which the individual response functions can be displayed and compared.

The relation in \cref{eq:bayes_ratio} also gives the connection to event reweighting. If the fitted logit were equal to the exact log density ratio, the weight that transforms the \pythia{} distribution into the \herwig{} distribution would be
\begin{equation}
	w_s(x)
	=
	\frac{p_H(x)}{p_P(x)}
	=
	\exp[\ell_s(x)].
	\label{eq:event_weight_general}
\end{equation}
For an arbitrary integrable observable $O(x)$, importance sampling then gives
\begin{align}
	\mathbb{E}_{H}[O]
	&=
	\int \dd x\,p_H(x)\,O(x)
	\nonumber\\
	&=
	\int \dd x\,p_P(x)
	\frac{p_H(x)}{p_P(x)}\,O(x).
	\label{eq:importance_sampling_exact}
\end{align}
For a finite sample of \pythia{} events this expectation can be estimated with the self-normalized weighted average
\begin{equation}
	\mathbb{E}_{H}[O]
	\simeq
	\frac{
		\sum_{e\in P} w(x_e)\,O(x_e)
	}{
		\sum_{e\in P} w(x_e)
	}.
	\label{eq:importance_sampling}
\end{equation}
The interpretation is simple: a \pythia{} event receives a large weight in a region of observable space that is relatively more populated by \herwig{}, and a smaller weight where the opposite is true.

This procedure requires sufficient overlap between the two distributions. In particular, regions populated by the target \herwig{} distribution must also be represented in the source \pythia{} sample. A reweighting procedure cannot create events in a region where the source sample has no support. Even before this extreme limit is reached, a very small number of events can acquire most of the total weight. The resulting weighted sample then carries little statistical information. This support issue will become important in Analysis II in \cref{sec:analysisII}.

For the additive KAN, inserting \cref{eq:add_kan} into \cref{eq:event_weight_general} gives
\begin{equation}
	w_{s,\mathrm{add}}(x)
	=
	e^{b_s}
	\prod_{i=1}^{8}
	e^{f_{s,i}(x_i)}.
	\label{eq:weight_factorization}
\end{equation}
We can therefore associate with each observable the factor
\begin{equation}
	w_{s,i}(x_i)
	=
	e^{f_{s,i}(x_i)}.
	\label{eq:component_weight}
\end{equation}
The full product in \cref{eq:weight_factorization} is an exact factorization of the \emph{fitted additive KAN weight}. An individual factor $w_{s,i}(x_i)$ should not, however, be identified with the one-dimensional marginal density ratio $p_H(x_i)/p_P(x_i)$.The additive functions are obtained from a joint fit to all eight observables, so their decomposition is basis dependent, while much like the choice of a gauge, the centering convention only fixes the otherwise arbitrary additive offsets of the individual components. The individual factors are therefore components of the fitted multivariate reweighting function. Whether one of them can be interpreted usefully on its own must be established from its statistical support and from an explicit reweighting test.

\subsection{Training methodology}
\label{sec:training}

Going back to the data split introduced in \cref{sec:cohort}, we now describe in more detail how the training, validation, and testing samples are constructed. There are two successive levels of data separation. First, the common hard-event cohort is divided into a \emph{development sample} and a disjoint \emph{final-test sample}. Within the development sample we use a five-fold cross-fitting procedure for training, validation, and out-of-fold evaluation. The final-test sample is kept separate throughout development and is used only to check whether the conclusions persist on held-out hard events.

The training procedure takes place entirely within the development sample. Each development hard event is assigned deterministically to one of five \emph{cross-validation folds}. The fold assignment is made using the hard-event identifier, so all representations originating from the same hard scattering remain together. Consider one of the five folds, labelled $k$. In that training cycle, fold $k$ is kept aside as the \emph{scoring fold}. A second fold is used as the \emph{validation fold} to determine the early-stopping point, while the remaining three folds are used to train the network. The trained model is then evaluated on fold $k$, which has participated neither in parameter optimization nor in the early-stopping decision.

\begin{algorithm}[tbph]
	\caption{Cross-fitted performance evaluation}
	\label{alg:crossfit}
	\DontPrintSemicolon
	
	\KwIn{Development sample
		$\mathcal{D}_{\rm dev}=\bigsqcup_{k=0}^{4}F_k$}
	\KwOut{Out-of-fold predictions and OOF AUC}
	
	\ForEach{generator stage $s$ and architecture
		$M\in\{[8,1],[8,4,1]\}$}{
		
		\For{$k=0,\ldots,4$}{
			$F_{\rm score}\leftarrow F_k$\;
			$F_{\rm val}\leftarrow F_{(k+1)\bmod 5}$\;
			$F_{\rm train}\leftarrow
			\mathcal{D}_{\rm dev}\setminus
			(F_{\rm score}\cup F_{\rm val})$\;
			
			\ForEach{random seed
				$\rho\in\mathcal{S}_{\rm seed}$,
				$|\mathcal{S}_{\rm seed}|=5$}{
				
				Standardize inputs using $F_{\rm train}$ only\;
				Train $M$ on $F_{\rm train}$ with early stopping on
				$F_{\rm val}$\;
				Evaluate the raw logit
				$\ell^{(\rho)}_{s,M}(x)$ on $F_{\rm score}$\;
			}
			
			Average the five raw logits on $F_{\rm score}$\;
		}
		
		Combine the five scoring folds to form the OOF predictions\;
		Compute the OOF AUC\;
	}
\end{algorithm}

The procedure in algorithm \ref{alg:crossfit} is repeated until each of the five folds has served once as the scoring fold. Every development event therefore receives a prediction from a network that was not trained or validated on that event. These are the \emph{out-of-fold} (OOF) predictions. For every training configuration we train five independently initialized networks, and their raw logits are averaged before evaluating the OOF performance. 

Both KAN architectures are implemented using the pyKAN package. The central additive model has width $[8,1]$, while the deeper diagnostic model has width $[8,4,1]$. Both use five spline-grid intervals and cubic splines. The networks are trained with binary cross-entropy acting on the raw logit. We use the Adam optimizer \cite{kingma2014adam} with a learning rate of $10^{-3}$, a batch size of $256$, and a maximum of $200$ epochs. The regularization strength is $\lambda=0.002$ with entropy coefficient $1.0$. Early stopping is determined exclusively from the validation fold.

A second set of fits is required for the functional analysis since the cross-fitted models above are used only to measure development-sample performance. Further, each scoring fold is evaluated by a different trained model, hence they do not define one common set of response functions over the development sample. We therefore fix one development fold for validation and train the additive KAN on the remaining four folds, repeating the fit for five random seeds. The five resulting sets of response functions are kept separate rather than averaged. Subset recomposition, Shapley allocation, and downstream transport are performed independently for each seed, and the resulting quantities are then summarized over the five fits. The exported functions are checked numerically against their corresponding live networks before being used in these analyses. The separate development fits used to construct the explicit response functions are summarized in algorithm \ref{alg:function_export}.

\begin{algorithm}[tbph]
	\caption{Training and export of additive KAN response functions}
	\label{alg:function_export}
	\DontPrintSemicolon
	
	\KwIn{Development sample $\mathcal{D}_{\rm dev}$}
	\KwOut{Centered functions $f^{(\rho)}_{s,i}$ and bias $b_s^{(\rho)}$}
	
	Fix a predetermined fold $F_{k_*}$ as $F_{\rm val}$\;
	$F_{\rm train}\leftarrow
	\mathcal{D}_{\rm dev}\setminus F_{\rm val}$\;
	Fit the input standardization using $F_{\rm train}$ only\;
	
	$X_{\rm ref}\leftarrow F_{\rm train}$,
	$\quad\bar{x}\leftarrow\langle x\rangle_{X_{\rm ref}}$\;
	
	\ForEach{generator stage $s$}{
		\ForEach{random seed $\rho\in\mathcal{S}_{\rm seed}$}{
			
			Train an additive $[8,1]$ KAN on $F_{\rm train}$
			with early stopping on $F_{\rm val}$\;
			
			$\ell_0\leftarrow
			\ell^{(\rho)}_{s,\mathrm{add}}(\bar{x})$\;
			
			\For{$i=1,\ldots,8$}{
				Vary $x_i=z$ over its $F_{\rm train}$ support,
				with $x_{j\neq i}=\bar{x}_j$\;
				
				$\widetilde f^{(\rho)}_{s,i}(z)
				\leftarrow
				\ell^{(\rho)}_{s,\mathrm{add}}
				(\bar{x}_1,\ldots,z,\ldots,\bar{x}_8)-\ell_0$\;
				
				$c^{(\rho)}_{s,i}
				\leftarrow
				\langle\widetilde f^{(\rho)}_{s,i}(x_i)
				\rangle_{X_{\rm ref}}$\;
				
				$f^{(\rho)}_{s,i}(z)
				\leftarrow
				\widetilde f^{(\rho)}_{s,i}(z)-c^{(\rho)}_{s,i}$\;
			}
			
			$b_s^{(\rho)}
			\leftarrow
			\ell_0+\sum_i c^{(\rho)}_{s,i}$\;
			
			Export $b_s^{(\rho)}$ and
			$\{f^{(\rho)}_{s,i}\}_{i=1}^{8}$\;
			Verify numerical agreement with the trained network\;
		}
	}
\end{algorithm}

It is therefore useful to distinguish the two roles clearly. The \emph{cross-fitted models} provide OOF predictions and are used to quote development-sample classification performance. The \emph{development/export models} provide the explicit functions $f_{s,i}(x_i)$ that form the functional anatomy studied in the remainder of the paper. Both are constructed entirely without access to the final-test sample. 

Finally, a deeper KAN is used only to test whether the deliberately additive architecture leaves substantial generator-discriminating information unmodelled. Its output is denoted by $\ell_{s,\mathrm{deep}}(x)$, while the additive output is $\ell_{s,\mathrm{add}}(x)$. We compare their OOF AUCs to quantify the loss of discriminating power associated with imposing the additive form. The pointwise difference
\begin{equation}
	\ell_{s,\mathrm{deep}}(x)
	-
	\ell_{s,\mathrm{add}}(x)
\end{equation}
is not interpreted as a unique interaction contribution. The two networks are trained independently, so this difference is used only as a diagnostic of non-additive information that the central $[8,1]$ model may not have captured.

\subsection{Exact subset recomposition and Shapley allocation}
\label{sec:shapley_intro}

The additive structure in \cref{eq:add_kan} allows us to ask which combinations of observables carry the generator separation without training a new network for every combination. Let
\begin{equation}
	\mathcal{O}=\{1,\ldots,8\}
\end{equation}
denote the complete set of eight observables. For any subset $S\subseteq \mathcal{O}$, we construct
\begin{equation}
	\ell_{s,S}(x)
	=
	b_s+\sum_{i\in S}f_{s,i}(x_i).
	\label{eq:subset_logit_main}
\end{equation}
Note that no network is retrained in this operation. We simply retain the functions belonging to the chosen observables and remove the others from the same fitted additive KAN. With eight observables there are
\begin{equation}
	2^8=256
\end{equation}
possible subsets. The empty subset contains only the constant $b_s$, while the full subset reproduces the complete additive KAN. We can therefore study the entire subset space exactly within the fitted model.

The only remaining question is then, how to assign the performance of these different subsets to the individual observables. An observable can appear unimportant when considered alone but become useful when combined with another observable, or vice versa. Ranking only the eight single-observable models would therefore depend strongly on which other information had already been included. We address this using Shapley values \cite{Shapley:1953}.

Suppose that $v(S)$ assigns a numerical value to every subset $S$. The Shapley value of observable $i$ is then
\begin{equation}
	\phi_i[v]
	=
	\sum_{S\subseteq \mathcal{O}\setminus\{i\}}
	\frac{|S|!\,(|\mathcal{O}|-|S|-1)!}{|\mathcal{O}|!}
	\left[
	v(S\cup\{i\})-v(S)
	\right].
	\label{eq:shapley_main}
\end{equation}
The quantity in square brackets measures how much the value changes when observable $i$ is added to a particular set of already available observables. The factorial factor averages this change over all possible orders in which the eight observables could have been introduced. The resulting Shapley value therefore measures the average marginal contribution of observable $i$ to the chosen quantity\footnote{For example, with only two observables $a$ and $b$,
	\begin{equation}
    	\phi_a
		=
		\frac{1}{2}
		\left[v(\{a\})-v(\varnothing)\right]
		+
		\frac{1}{2}
		\left[v(\{a,b\})-v(\{b\})\right].
	\end{equation}
	The first term asks what $a$ contributes when it is introduced before $b$. The second asks what it contributes when $b$ is already present. This simple example illustrates why the Shapley construction is more informative than ranking observables one by one.} $v$.

Since only the amount of generator separability is relevant for the subset
allocation, we remove the arbitrary orientation of the classifier score by
using the direction-folded ROC AUC,
\begin{equation}
	A_{\rm fold}
	=
	\max(A,1-A)
	=
	\frac{1}{2}
	+
	\left|A-\frac{1}{2}\right|,
	\label{eq:folded_auc}
\end{equation}
where $A$ is the ordinary ROC AUC. Thus $A_{\rm fold}=0.5$ corresponds to no
separation, independently of whether a particular recomposed score is oriented
toward \pythia{} or \herwig{}.

We use two different choices of the value function. The first measures \emph{generator separation}. For a subset $S$, we evaluate the recomposed logit $\ell_{s,S}$ on \pythia{} and \herwig{} events and define
\begin{equation}
	v_{\rm sep}(S)
	=
	A_{\rm fold}\!\left[\ell_{s,S}\right].
\end{equation}
Here an AUC of $0.5$ corresponds to no generator separation and an AUC of unity to complete separation. The associated Shapley values quantify how the separating power of the fitted additive KAN is distributed among the eight observables.

The second value function measures \emph{reweighting closure}. An independent classifier, referred to below as the judge, measures the residual separation between \pythia{} and \herwig{}. For each subset we reweight the \pythia{} events by $\exp[\ell_{s,S}(x)]$ and ask how much the judge's separation is reduced. Schematically,
\begin{equation}
	v_{\rm cl}(S)
	=
	A_{\rm fold,judge}^{\rm unweighted}
	-
	A_{\rm fold,judge}^{\rm weighted,S}.
\end{equation}
A positive value therefore means that the selected subset moves the weighted \pythia{} sample toward \herwig{} according to this independent classifier. The closure Shapley values should be interpreted as an allocation of the closure diagnostic within the fitted additive model. They do not by themselves establish that an individual component defines a statistically reliable importance weight. The support of such a factor is assessed separately through its effective sample size and influence diagnostics in Analysis II. This distinction is particularly relevant for the stage A mass components, which enter the functional subset decomposition but are not interpreted as supported standalone transport weights.

The separation and closure constructions answer related but different questions. The first asks which observables contain generator-discriminating information. The second asks which parts of the learned reweighting function actually reduce the residual generator difference. The Shapley values satisfy the exact sum rule
\begin{equation}
	\sum_{i\in \mathcal{O}}\phi_i[v]
	=
	v(\mathcal{O})-v(\varnothing),
\end{equation}
so the total change between the empty and complete additive model is fully distributed among the eight observables. Statistical intervals are obtained by resampling complete hard events, so that the \pythia{} and \herwig{} representations originating from the same hard scattering are always resampled together. Further details for this construction procedure are given in \cref{app:shapley}.

Finally, the Shapley allocations should be interpreted within the scope of the present model. They quantify how the chosen eight observables contribute to the separation or closure of the fitted additive KAN. They are not causal fractions of parton-shower, hadronization, or MPI physics. In Analysis I, the physical interpretation comes from comparing the observable-level structures and their relative importance across the three controlled generator stages. In Analysis II, we go one step further and test whether selected Stage A factors retain their reweighting power when carried downstream through hadronization and the full-generator evolution.

\section{Analysis I: stage-dependent anatomy of the generator difference}
\label{sec:analysisI}

\begin{figure}[tbph]
	\centering
	\includegraphics[width=0.72\textwidth]{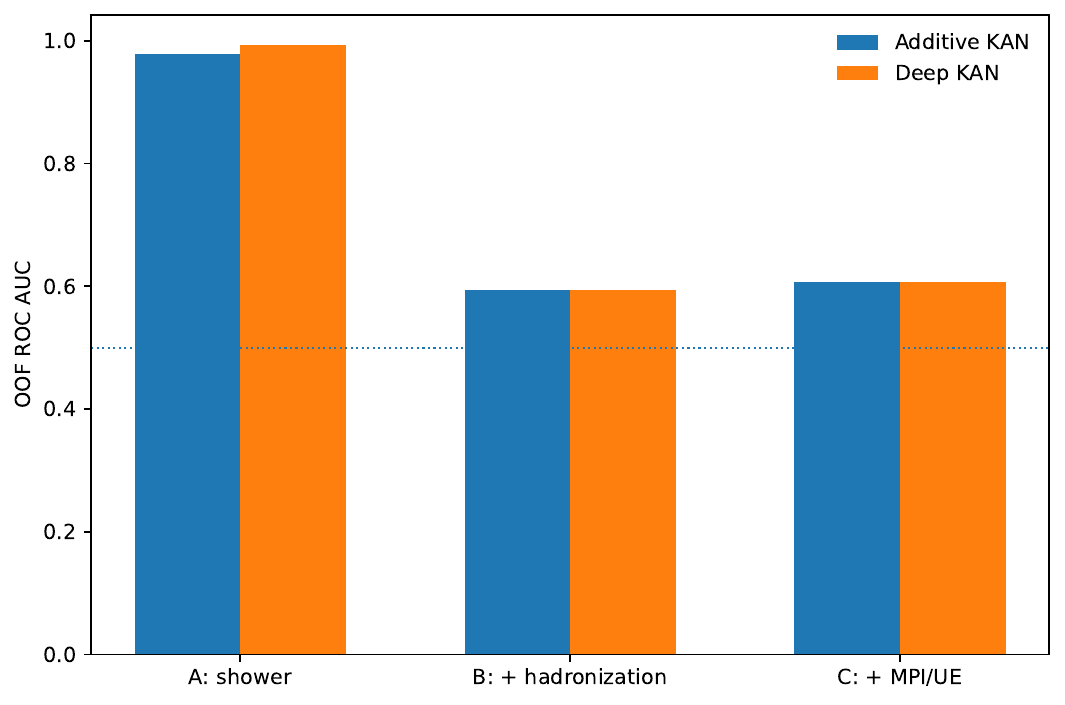}
	\caption{Out-of-fold \pythia{}-\herwig{} separation on the common eight-observable representation. The dotted reference corresponds to chance-level separation.}
	\label{fig:staged_auc}
\end{figure}

The first analysis asks how strongly \pythia{} and \herwig{} differ when the hard-event cohort and the eight-observable representation are held fixed. The main results are illustrated by \cref{fig:staged_auc,tab:staged_auc} showing the OOF development-sample separation.
The stage A representation makes the two generators almost perfectly distinguishable. Once hadronization is enabled the separation collapses to a modest level, and the full generator produces only a partial increase. Because the same hard scatterings are being compared, this pattern cannot be attributed to different matrix-element kinematics or a different selected event population.
\begin{table}[tbph]
	\centering
	\caption{OOF generator separation on the development sample. The final column diagnoses additional non-additive discriminating information.}
	\label{tab:staged_auc}
	\begin{tabular}{lccc}
		\toprule
		Generator configuration & additive KAN & deep KAN & $\Delta\auc$ \\
		\midrule
		A: shower only & \AAddA & \ADeepA & +\ADeltaA \\
		B: shower + hadronization & \AAddB & \ADeepB & +\ADeltaB \\
		C: full generator & \AAddC & \ADeepC & +\ADeltaC \\
		\bottomrule
	\end{tabular}
\end{table}
For the functional analysis, the more interesting observation is that the additive KAN retains almost all of the separation available to the deeper model. The deep network gains only $\ADeltaA$ in AUC at stage A and essentially nothing at B or C. One should be careful here and do not na\"ively assume that this proves that the true density ratio factorizes or that the inputs are uncorrelated. It actually establishes an empirical statement about the chosen representation that forcing the logit into a sum of explicit one-dimensional functions sacrifices only little of the observed generator information. We can therefore study the individual response functions while retaining nearly all of the discriminating information captured by the deeper KAN in this eight-observable representation.

\subsection{Response functions across stages}
\label{sec:response_reading}

We first examine the one-dimensional response functions learned by the additive KAN at the three generator stages. The purpose of these plots is not only to identify which observable produces a large classifier response, but also to determine whether that response lies in a region of observable space that is sufficiently populated to support a meaningful reweighting interpretation.

\begin{figure}[p]
	\centering
	\includegraphics[
	width=0.98\textwidth,
	trim=0 0 0 50,
	clip
	]{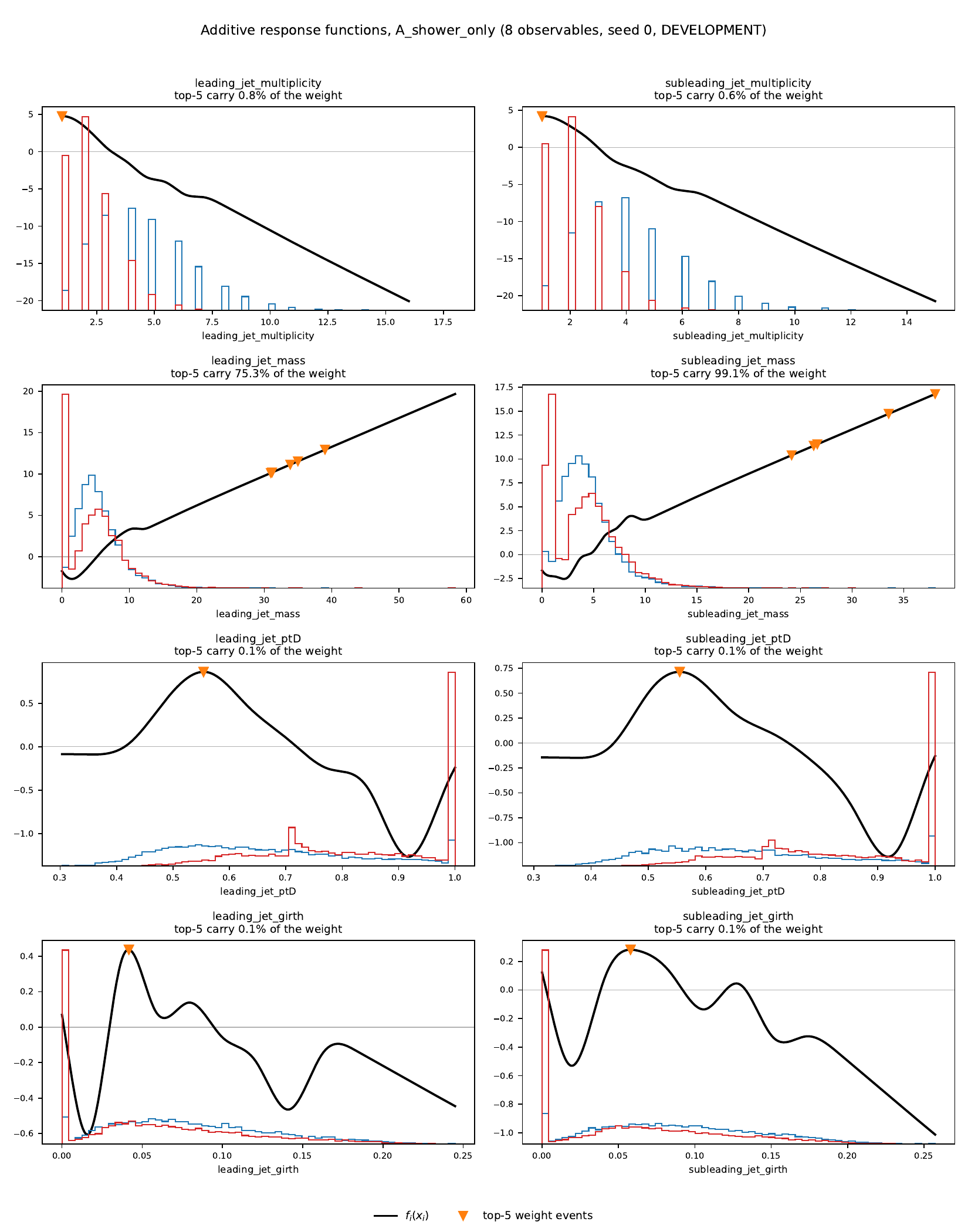}
	\caption{Representative additive-KAN response functions at stage A.}
	\label{fig:responseA}
\end{figure}

\begin{figure}[p]
	\centering
	\includegraphics[
	width=0.98\textwidth,
	trim=0 0 0 50,
	clip
	]{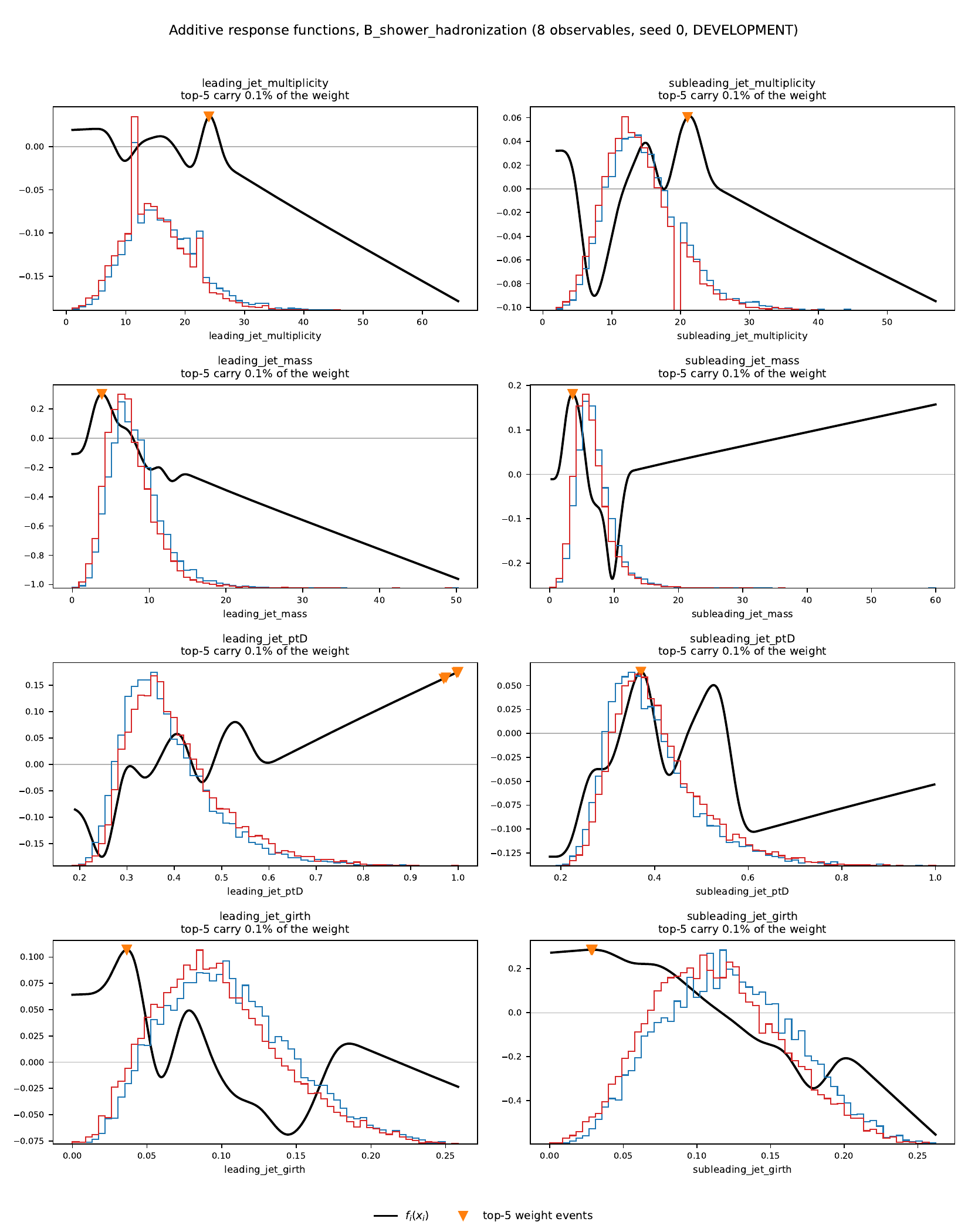}
	\caption{Representative additive-KAN response functions at stage B.}
	\label{fig:responseB}
\end{figure}

\begin{figure}[p]
	\centering
	\includegraphics[
	width=0.98\textwidth,
	trim=0 0 0 50,
	clip
	]{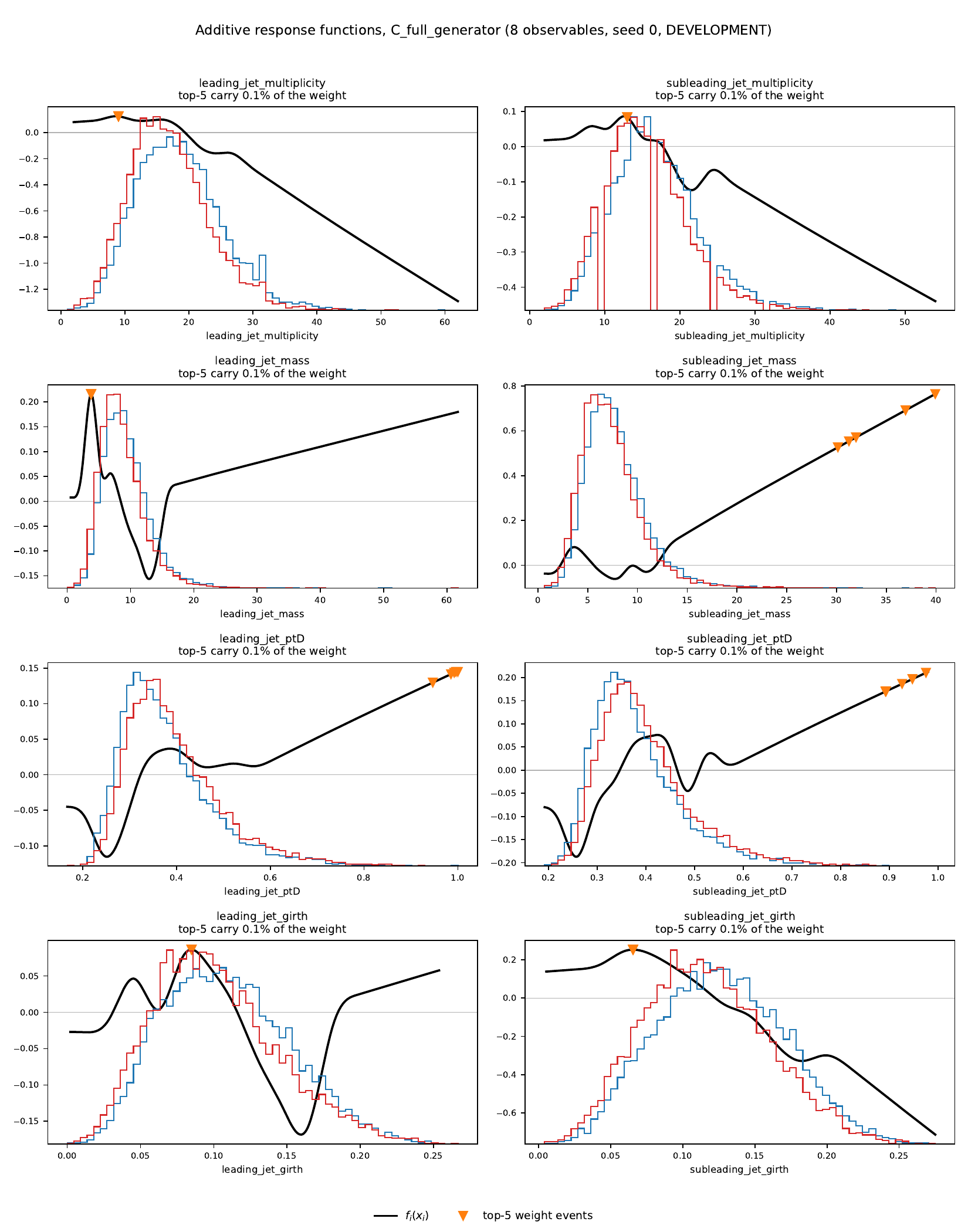}
	\caption{Representative additive-KAN response functions at stage C.}
	\label{fig:responseC}
\end{figure}

\Cref{fig:responseA,fig:responseB,fig:responseC} show representative additive KANs at stages A, B, and C, respectively. Each figure is taken from one of the five additive models trained on the development sample. The final-test sample is not used here, since the response functions and the choice of components to be studied downstream are fixed before that sample is \emph{opened}.

Each panel corresponds to one of the eight input observables. The black curve shows the centered KAN component $f_{s,i}(x_i)$ defined in \cref{eq:add_kan}. Since \pythia{} is labelled as $0$ and \herwig{} as $1$, a positive value of $f_{s,i}$ increases the fitted logit in the \herwig{}
direction, while a negative value increases it in the \pythia{} direction. The corresponding multiplicative component of the fitted \pythia{}$\to$\herwig{} event weight is
\begin{equation}
	w_{s,i}(x_i)=\exp[f_{s,i}(x_i)].
\end{equation}

The blue and red step distributions show the locations populated by \pythia{} and \herwig{}, respectively. Their vertical normalization is rescaled for display and should therefore not be compared directly with the vertical scale of $f_{s,i}$.  Their role is instead to show where the learned response is supported by events. This distinction is particularly important for reweighting from \pythia{} to \herwig{}, for which the \pythia{} distribution defines the available source support.

The orange markers provide a complementary diagnostic. For each component they indicate the positions of the five \pythia{} events carrying the largest values of $w_{s,i}$.  The number quoted above each panel gives the fraction of the total normalized component weight carried by these five events. A large response in a sparsely populated region is therefore not automatically
evidence for a robust generator difference. If a few source events carry a large fraction of the weight, the corresponding reweighting direction has poor effective statistical support even though the fitted function itself is perfectly well defined. Several of the five highest-weight events can have the same value of a given observable, particularly for the discrete constituent multiplicities. Their markers then overlap, so fewer than five orange triangles may be visually distinguishable even though the quoted weight fraction always refers to all five events.

At stage A, the dominant structure is immediately visible in the constituent multiplicities. Both multiplicity functions give a positive contribution to the \herwig{} logit at small multiplicity and become increasingly negative as the number of shower constituents increases. Importantly, this structure is supported by a sizeable population of events rather than by a few isolated configurations. The five largest component weights carry only a small fraction of the total multiplicity weight.

The jet-mass functions behave very differently. Their fitted responses grow strongly in regions where the \pythia{} source population is already extremely small. Consequently, a few events dominate the corresponding component weights. This is especially pronounced for the two mass observables. The functions therefore make the mass dependence visible, but the resulting standalone mass reweighting is statistically fragile, in the sense that the effective sample size becomes very small and the weighted estimate is controlled by only a few source events, making it highly sensitive to their presence or absence. The distinction between a visible learned response and a well-supported reweighting direction will be important in Analysis II.

The picture changes substantially at stage B, after hadronization is included in both generator chains. The very large shower-level multiplicity response is no longer present. Across the populated regions the individual functions are much smaller, consistent with the substantially weaker overall generator separation at this stage. At the same time, non-trivial dependence remains in the jet-mass and shape observables. Unlike the stage A mass response, these components do not exhibit an extreme concentration of their event weights in a handful of source events.

This observation should not be interpreted as isolating a pure ``hadronization effect''.  Stage B compares the two generators after each has evolved the common hard event through its own shower and its own hadronization model. The comparison therefore tells us how the \pythia{}-\herwig{} difference is reorganized when hadronization is introduced, rather than providing a causal decomposition of the two hadronization models.

At stage C, where MPI and the associated underlying-event activity are also included, a coherent multiplicity dependence becomes visible again. In both the leading and subleading jets, lower multiplicities contribute positively to the fitted \herwig{} logit while larger multiplicities contribute negatively. Mass and shape dependence remains present, but the response is no longer organized in the same way as at stage B. Some functions also grow in sparsely populated tails. Such regions are retained in the plots because they are part of the learned function, but they are not assigned physical importance unless they are supported by the event population and remain stable in the quantitative analysis.

The three stages therefore already suggest a qualitative reorganization of the generator difference. At stage A, constituent multiplicity provides the clearest well-supported response, while the large mass responses occur in regions with poor source statistics. The response functions at stages B and C show a different pattern, indicating that the relative importance of the observables is not preserved through the generator evolution. The response amplitudes alone, however, do not provide a quantitative measure of how much each observable contributes to the full generator separation. We therefore determine this hierarchy using the complete subset decomposition and Shapley allocation of the additive model.

\subsection{Shapley allocation across generator stages}
\label{sec:stage_anatomy}

The response functions provide a direct view of the learned observable dependence, but their amplitudes alone do not give a unique measure of how much each observable contributes to the full multivariate classifier. We therefore use the exact subset recomposition and Shapley construction as introduced in \cref{sec:shapley_intro}.

\begin{figure}[t]
	\centering
	\begin{subfigure}{0.49\textwidth}
		\centering
		\includegraphics[width=\textwidth]
		{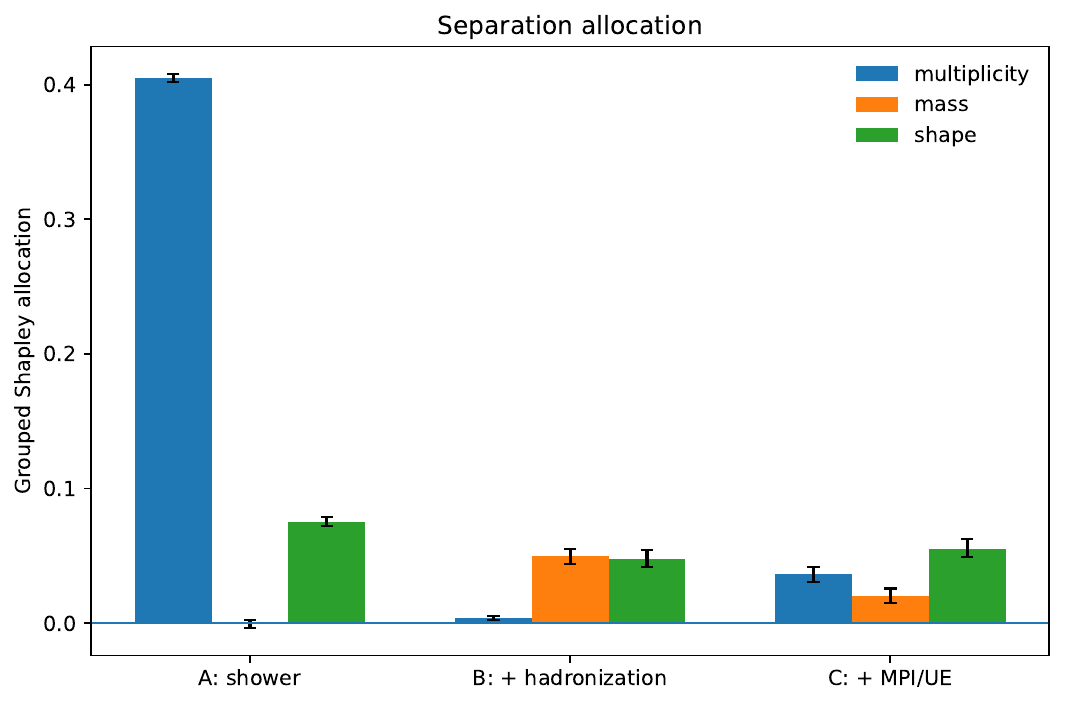}
		\caption{Separation.}
	\end{subfigure}\hfill
	\begin{subfigure}{0.49\textwidth}
		\centering
		\includegraphics[width=\textwidth]
		{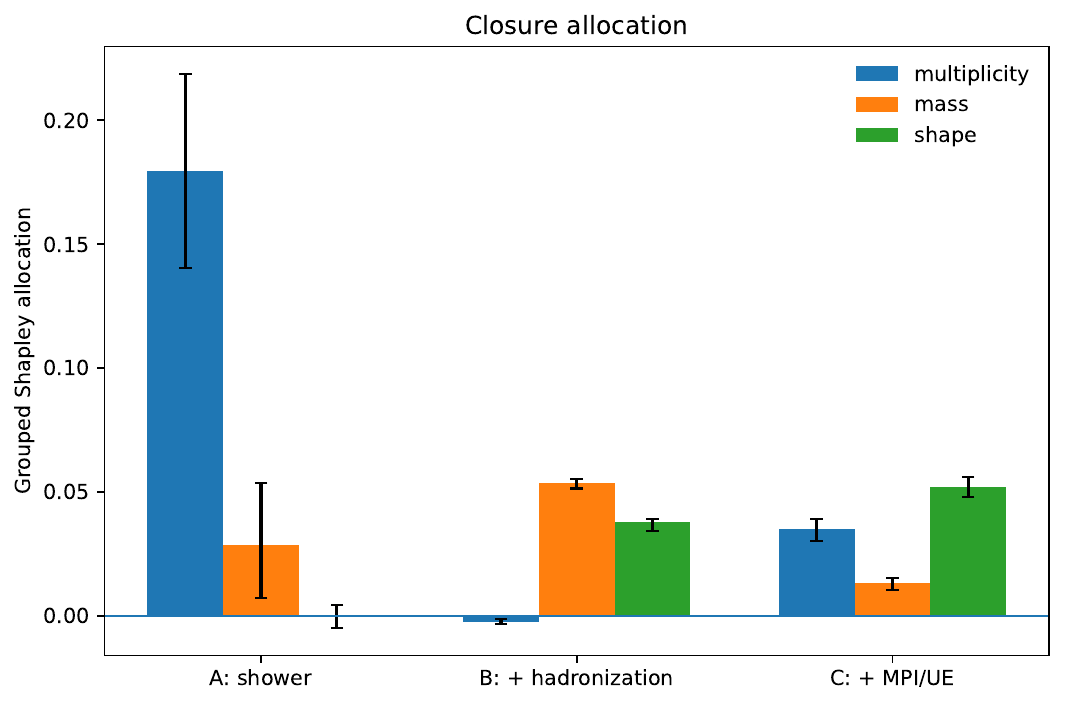}
		\caption{Reweighting closure.}
	\end{subfigure}
	\caption{Grouped Shapley allocations for the multiplicity, mass, and shape
		sectors at the three generator stages.  Error bars show hard-event-clustered
		bootstrap intervals.}
	\label{fig:shapley_sectors}
\end{figure}

For the purpose of the physics comparison, the eight observables are grouped into three sectors,
\begin{equation}
	\begin{aligned}
		\text{multiplicity}:&\quad (n_1,n_2),\\
		\text{mass}:&\quad (m_1,m_2),\\
		\text{shape}:&\quad
		(p_{T,1}^{D},g_1,p_{T,2}^{D},g_2).
	\end{aligned}
\end{equation}
We evaluate the two value functions defined previously. The \emph{separation} allocation measures how the generator-discriminating AUC of the fitted additive model is distributed among the observable sectors. The \emph{closure} allocation measures which sectors are most effective in reducing the residual \pythia{}-\herwig{} separation when used for reweighting. The corresponding grouped Shapley values are shown in \cref{fig:shapley_sectors}. Further details of the Shapley construction, statistical procedure and obtained values are given in \cref{app:shapley}.

The separation and closure allocations give the same qualitative evolution,
\begin{equation}
	\begin{aligned}
		A:&\quad \text{multiplicity dominated},\\
		B:&\quad \text{mass and shape dominated},\\
		C:&\quad \text{shape led, with a substantial multiplicity component}.
	\end{aligned}
	\label{eq:stage_anatomy}
\end{equation}
At stage A the multiplicity sector dominates both allocations by a large margin. The closure allocation also assigns a smaller positive contribution to mass but as discussed in Analysis II, this should not be interpreted as evidence that the corresponding mass factor defines a statistically reliable standalone importance weight. At stage B the multiplicity contribution becomes small, while mass and shape carry the residual generator difference. The two sectors are nearly equal in the separation allocation, while mass gives the larger contribution to reweighting closure. At stage C the structure changes again. Shape gives the largest contribution to both separation and closure, with a substantial but smaller multiplicity component and a still smaller mass contribution. The full-generator difference is therefore best described as a shape-led mixed structure rather than as a multiplicity-dominated one.

\section{Analysis II: transporting the shower-level functional response}
\label{sec:analysisII}

The common hard-event identity allows us to ask whether an observable-level difference learned at the shower stage remains relevant after the subsequent generator evolution. For a stage A component $i$, we evaluate the corresponding factor on the \pythia{} representation of hard event $h$,
\begin{equation}
	w_{A,i}(h)
	=
	\exp\!\left[
	f_{A,i}\!\left(x^{P}_{A,i}(h)\right)
	\right].
	\label{eq:transport_weight}
\end{equation}
This number is then attached to the same hard event at stages B and C. In particular, $f_{A,i}$ is never re-evaluated using $x_{B,i}$ or $x_{C,i}$. The transport therefore tests whether the information encoded by a shower-level response remains useful for reweighting the downstream event ensemble. The construction therefore transports information associated with the same
underlying hard scattering, rather than requiring a matched shower or jet trajectory between the stages.

At each target stage, a separate cross-fitted KAN classifier is trained on the unweighted \pythia{} and \herwig{} samples. The transported weights are then applied to the \pythia{} events when evaluating the AUC of this independent classifier. Since an AUC below $0.5$ simply corresponds to the same separation with the classifier direction reversed. As in \cref{eq:folded_auc}, we fold the AUC about $0.5$ and retain only the amount of residual generator separation. Thus $0.5$ denotes no separability, while values farther from $0.5$ indicate increasing separation. A successful transport should therefore move the folded AUC toward $0.5$, but this is interpreted together with the effective sample size and the stability checks described in appendix \ref{app:statistics}. A value close to $0.5$ by itself is not sufficient if the weighted sample is dominated by only a few events.

\subsection{Transport of the leading-multiplicity factor}
\label{sec:mult_transport}

\begin{figure}[t]
	\centering
	\begin{subfigure}{0.49\textwidth}
		\centering
		\includegraphics[width=\textwidth]{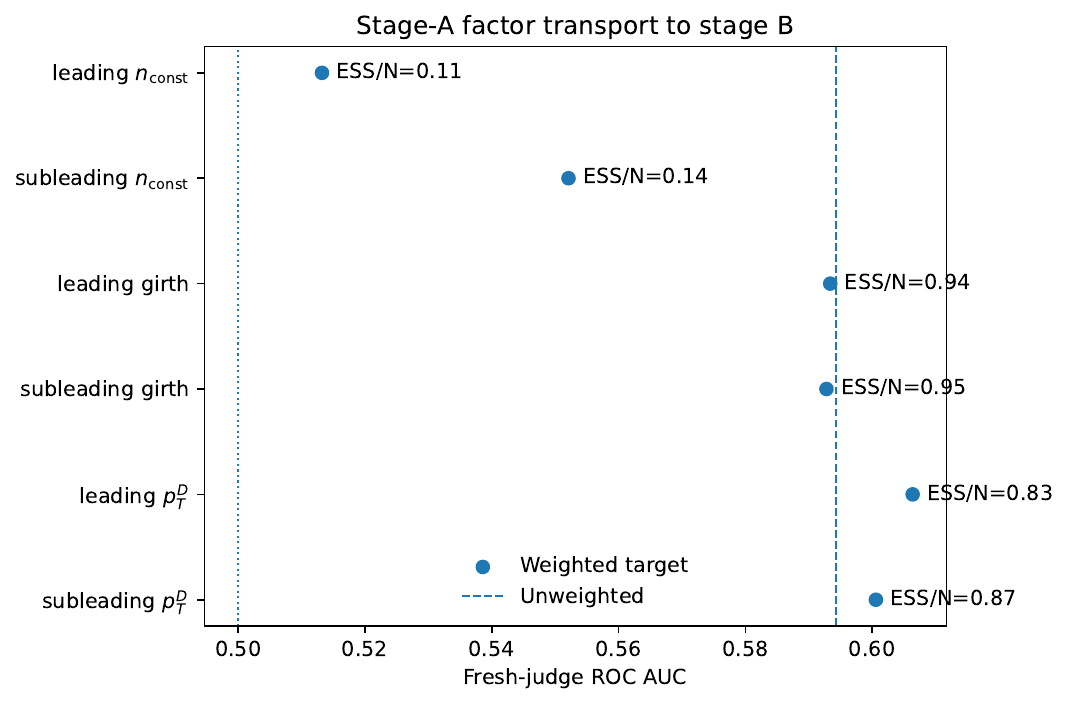}
		\caption{Transport to stage B.}
	\end{subfigure}\hfill
	\begin{subfigure}{0.49\textwidth}
		\centering
		\includegraphics[width=\textwidth]{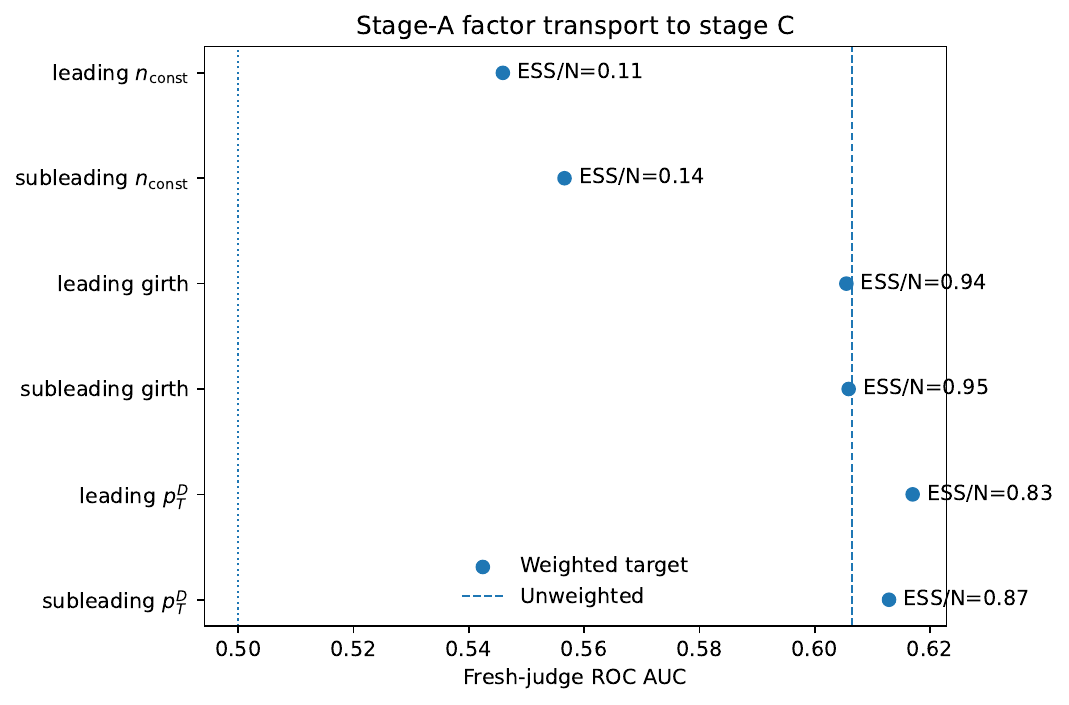}
		\caption{Transport to stage C.}
	\end{subfigure}
	\caption{Transport of the individual stage A factors to the later generator
		stages. The dashed line shows the unweighted generator separation and
		the dotted line corresponds to an AUC of $0.5$. The mass factors are
		omitted because their effective sample sizes are too small for a reliable
		weighted comparison.}
	\label{fig:factor_transport}
\end{figure}

The clearest transport is obtained from the stage A leading-jet multiplicity factor. On the OOF development sample, the target-stage AUC changes as\footnote{These unweighted reference AUCs need not coincide exactly with the additive-KAN values quoted in \cref{tab:staged_auc}. The latter are the OOF AUCs of the stage-specific additive KANs used for the functional decomposition, whereas the transport analysis is evaluated with a separate cross-fitted target-stage KAN judge.}
\begin{equation}
	B:\quad \DevBUnw\longrightarrow\DevBLeadMult,
	\qquad
	C:\quad \DevCUnw\longrightarrow\DevCLeadMult,
	\label{eq:mult_transport_dev}
\end{equation}
with $\ess/N=\LeadMultESSDev$, where $\ess$ is the weight-based effective sample size defined in \eqref{eq:kish} for the reweighted \pythia{} sample and $N$ is the original number of source events. The ratio $\ess/N$ therefore quantifies how much effective statistical support remains after reweighting. The quoted values are the means over the five independently trained stage A models, and all five give the same transport direction. For each source seed, the same stage A weight vector is carried unchanged to both target stages, so its effective-sample fraction is also unchanged between B and C. The transport behaviour of the individual stage A factors is summarized in \cref{fig:factor_transport}.

\begin{figure}[t]
	\centering
	\includegraphics[width=0.82\textwidth]{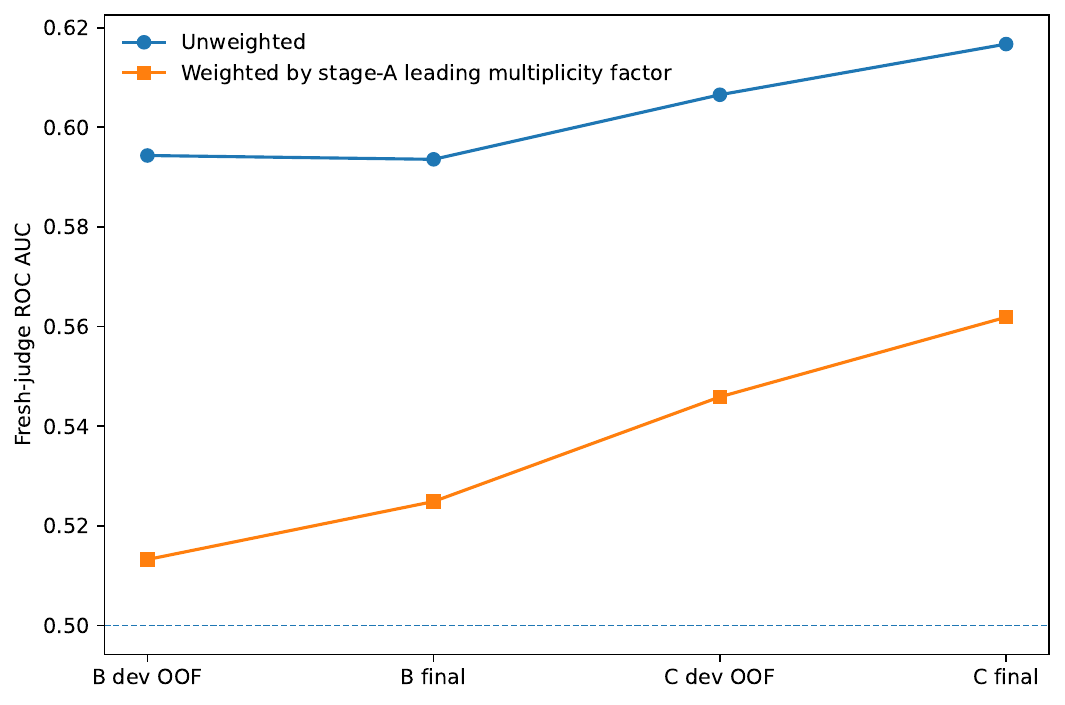}
	\caption{Development and final-test results for the transported stage A
		leading-multiplicity factor.}
	\label{fig:locked_transport}
\end{figure}

Applying the same five frozen stage A responses to the held-out final-test sample gives
\begin{equation}
	\begin{split}
		B:&\quad \TestBUnw\longrightarrow\TestBLeadMult,
		\qquad \ess/N=\LeadMultESSTest,\\
		C:&\quad \TestCUnw\longrightarrow\TestCLeadMult,
		\qquad \ess/N=\LeadMultESSTest.
	\end{split}
	\label{eq:locked_transport}
\end{equation}
The held-out final-test sample shows the same direction of transport for all five frozen source models. The reduction in residual generator separation is somewhat weaker than on the development sample, but the qualitative behaviour persists independently of the events used to identify the effect. The held-out final-test sample therefore provides an independent consistency check of the transport identified on the development sample, as shown in \cref{fig:locked_transport}.

The transport does not require the target-stage multiplicity itself to remain the dominant discriminator. The weight is determined from the shower-level event, while the same hard event can evolve into different masses, shapes, and correlated structures at later stages. The test therefore probes persistence of information identified at stage A, rather than persistence of the same one-dimensional observable distribution.

\subsection{Subleading multiplicity and shape transport}
\label{sec:null_transport}

The stage A subleading jet multiplicity provides a second statistically supported transport direction. On the development sample it changes the target-stage AUC as
\begin{equation}
	B:\quad \DevBUnw\longrightarrow\SubMultB,
	\qquad
	C:\quad \DevCUnw\longrightarrow\SubMultC,
\end{equation}
with $\ess/N=\SubMultESS$, and the transport direction is supported by all five stage A models. Interestingly, its downstream effect becomes relatively stronger at stage C, suggesting that
the shower-level multiplicity information carried by the two jets is reorganized as hadronization and the additional full-generator activity are introduced. We keep this result as an additional development-sample observation, since the leading-multiplicity factor was selected for the
final-test study.

The four shape factors behave differently. Their weights remain statistically well supported, with $\ess/N\simeq0.83$-$0.95$, yet none appreciably reduces the downstream generator separation. The leading $p_T^D$ factor gives AUCs $\LeadPTDB$ and $\LeadPTDC$, the leading-girth factor gives $\LeadGirthB$ and $\LeadGirthC$, the subleading $p_T^D$ factor gives $\SubPTDB$ and $\SubPTDC$, and the subleading-girth factor gives $\SubGirthB$ and
$\SubGirthC$. The absence of transport is therefore not a consequence of weight concentration.

This provides a useful physical distinction. Hadronization and the additional soft activity present at stage C can reorganize this internal structure, so a shape discrepancy identified at the shower level need not remain the appropriate downstream reweighting direction. This is particularly clear for subleading girth. It becomes important in the later within-stage decomposition while its stage A response has essentially no transport power. Thus, it shows that downstream importance does not necessarily imply upstream persistence.

\subsection{Mass factors and weight concentration}
\label{sec:mass_transport}

The mass factors encounter a different limitation. Their stage A response functions develop large values in sparsely populated mass tails, which become strongly concentrated importance weights after exponentiation as can be seen from \cref{fig:responseA}. For the leading and subleading jet masses,
\begin{equation}
	\frac{\ess}{N}=\LeadMassESS,
	\qquad
	\frac{\ess}{N}=\SubMassESS,
	\label{eq:mass_ess}
\end{equation}
respectively. The corresponding weighted AUCs are therefore not interpreted as physical transport measurements.

\begin{figure}[t]
	\centering
	\begin{subfigure}{0.49\textwidth}
		\centering
		\includegraphics[width=\textwidth]{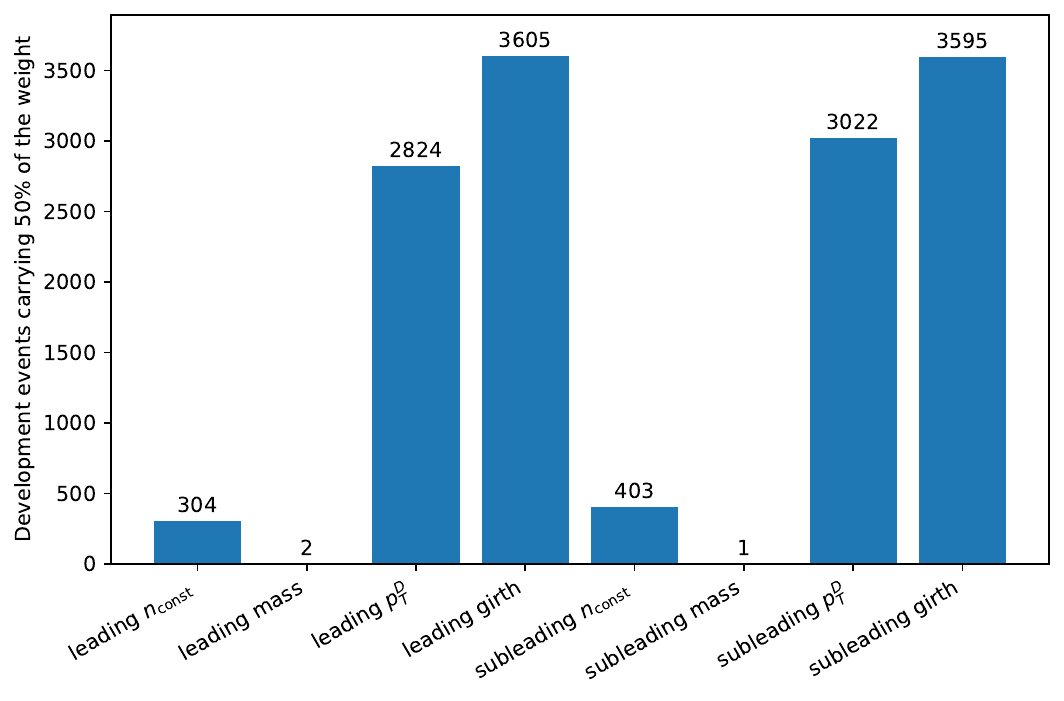}
		\caption{Weight concentration.}
	\end{subfigure}\hfill
	\begin{subfigure}{0.49\textwidth}
		\centering
		\includegraphics[width=\textwidth]{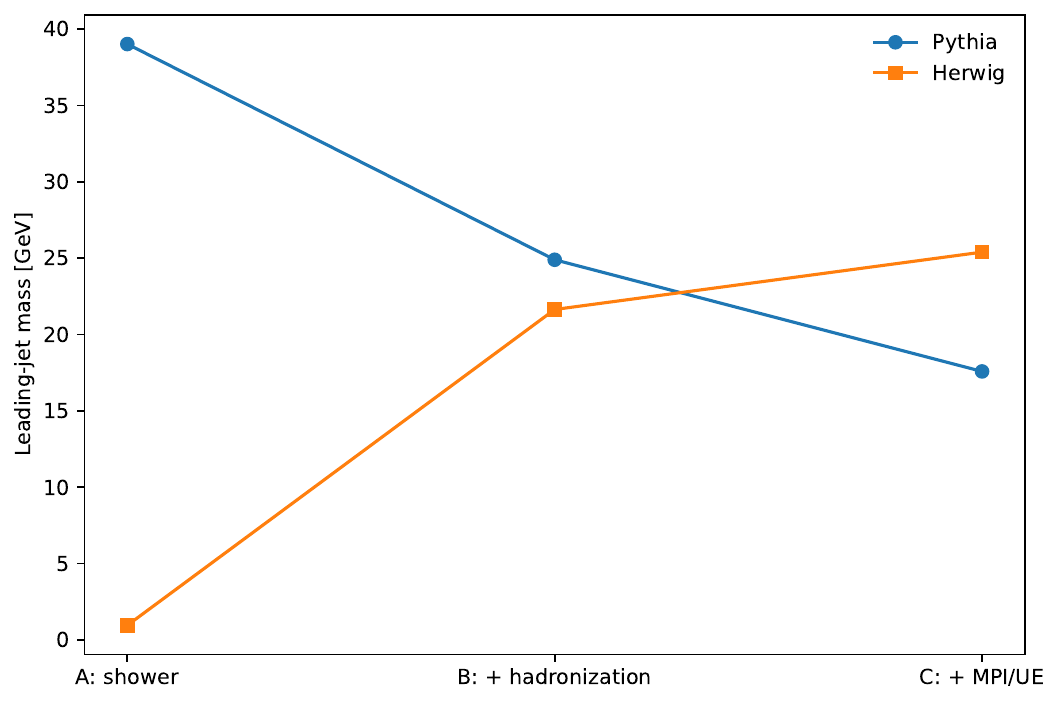}
		\caption{Dominant leading-mass hard event.}
	\end{subfigure}
	\caption{Statistical support of the stage A component weights. Left: number
		of development events required to accumulate $50\%$ of the normalized
		component weight. Right: leading-jet mass associated with the same
		influential hard event in the paired \pythia{} and \herwig{}
		configurations at stages A, B, and C.}
	\label{fig:weight_concentration}
\end{figure}

The severity of the concentration is shown in \cref{fig:weight_concentration}. After ordering the source events by their normalized component weight, only
\LeadMassHalfWeightEventsSeedZero{} events are required to accumulate $50\%$ of the total leading-mass weight, while only \SubMassHalfWeightEventsSeedZero{} event is required for subleading mass. In the representative fit shown here (for a particular seed),  half of the reweighted measure is controlled by only a very small number of rare source events. This is qualitatively different from the multiplicity and shape factors, whose weights are distributed over much broader sets of source events.

This behaviour originates from the source-stage tails. Exponentiating a large positive response in a region populated by very few \pythia{} events assigns those events a disproportionate fraction of the total weight. The mass transport is therefore limited by statistical support, rather than providing evidence for either persistence or disappearance of the stage A mass difference. Further stability tests are given in appendix \ref{app:weights}.

\section{Conclusion and discussion}
\label{sec:conclusion}

In this work, we have developed a KAN-based framework for studying generator-model differences as explicit functional structures, rather than only as differences between final-state distributions or global classifier scores. The central advantage of this framework is that the learned \pythia{}-\herwig{} discrepancy can be decomposed into explicit one-dimensional response functions that can be inspected, recomposed, stress-tested, and transported individually through different stages of event generation. This makes it possible not only to identify which observable directions carry a generator difference, but also to ask whether those directions remain meaningful as the event evolves from the shower to hadronization and to the full-generator description. In this sense, the framework provides a natural basis for future studies aimed at tracing and comparing generator-model dependence at the level of physically interpretable observable structures.

We demonstrate this approach using the same hard dijet events evolved through shower-only, shower with hadronization, and full-generator configurations of \pythia{} and \herwig{}, with the comparison performed on a common eight-observable dijet representation. The additive KAN is used here not because of any assumed advantage in classification performance, but because its functional form makes the learned discrepancy directly accessible. The small additional separation obtained with the deeper KAN shows that, within the chosen observable basis, the additive representation retains nearly all of the available discriminating information captured by the deeper KAN, while providing the explicit functional structure required for the subsequent analysis. Our main observations from the two analyses are:

\begin{description}
	\item[Stage-dependent anatomy:] Analysis I shows that the observable content of the generator difference changes substantially through the simulation chain. At the shower-only stage the difference is overwhelmingly associated with constituent multiplicity. After hadronization the overall \pythia{}-\herwig{} separation drops sharply and the remaining information is carried primarily by jet mass and shape. In the full-generator configuration, shape gives the largest allocation, while a substantial multiplicity component reappears. The resulting discrepancy is therefore a mixed shape-multiplicity structure rather than one controlled by a single observable sector. These changes should not be interpreted as a microscopic causal decomposition of showering, hadronization, and MPI, rather they instead establish how the observable representation of the \pythia{}-\herwig{} difference is reorganized as increasingly complete generator physics is included.
	
	\item[Transport and persistence:] Analysis II asks a different question: whether information learned at the shower level remains useful for reweighting the same hard events after their subsequent evolution. The shower-level multiplicity factors provide statistically supported downstream directions on the development sample, and the leading-multiplicity result shows the same behaviour on the held-out final-test sample. The shape factors give the complementary result. Their weights remain well supported statistically, yet their stage A responses do not reduce the downstream generator separation, even though shape becomes important again when the later stages are analysed directly. The downstream shape difference is therefore not simply the persistence of the shower-level shape response. In this sense, downstream importance does not imply upstream persistence.
	
	The mass factors reveal a third possibility. Their response functions are perfectly explicit, but the corresponding importance weights are dominated by sparsely populated stage A tails, leaving an extremely small effective sample. Their transport cannot therefore be interpreted reliably at the population level. This distinction is important for classifier-derived reweighting more generally. An explicit and reproducible learned function need not define a statistically usable weight. The same functional transparency that identifies a potentially interesting direction also exposes when the available sample does not support its interpretation.
\end{description}

There are some clear limitations of this work as well, which motivate future studies in this direction. We study one leading-order dijet process, one setup of each generator, and a compact set of eight high-level observables. The stage A jets are constructed from generator-internal shower partons and are therefore not directly measurable objects. Moreover, the density-ratio interpretation is conditional on the six-way common hard-event cohort. The differences in generator acceptance outside that cohort are not part of the learned weight. The numerical response functions should consequently not be regarded as universal calibration factors, and a two-generator comparison does not by itself define a complete theory uncertainty. Finally, the staged construction localizes where the observable structure changes, but does not isolate individual microscopic mechanisms such as hadronization, color reconnection, MPI, or shower recoil prescriptions.

At this point, It is also important to clarify the scope of the generator comparison itself. Throughout this work, \pythia{} and \herwig{} are treated as two fully specified generator configurations, including their respective shower, hadronization, MPI, and tune or default-parameter choices. We do not attempt to tune-match the two generators or to attribute an observed difference to any single parameter or microscopic modeling ingredient. Consequently, for example, the structure observed at stage C should be interpreted as the functional organization of the difference between the specified full-generator configurations, rather than as an isolated measurement of MPI, underlying-event, or tune effects. The same qualification also applies at stage A. Constituent multiplicity is used as a generator-level diagnostic under the native shower definitions, and the present analysis does not attempt to separate possible contributions from shower cutoffs, recoil prescriptions, or splitting implementations. Likewise, the transport construction follows information through the common hard-event identity. It tests whether a response learned at an earlier stage remains predictive downstream, rather than implying the propagation of an identical shower history or reconstructed jet through the different stages. These restrictions define the estimand of the present study rather than ambiguities in it, our aim is to characterize and trace the observable-level structure of the discrepancy between two reproducibly specified generator descriptions through a KAN-based framework.

Therefore, the natural next step is to apply this same KAN-based-framework to a broader set of processes, generator variations, and event representations while retaining the same support and closure diagnostics. One particularly interesting extension would be towards dark-sector searches. The Hidden-Valley implementations available in \pythia{} and, more recently, in \herwig{} make it possible to study strongly interacting dark sectors and dark-shower signatures, including semi-visible and emerging jets, within two genuinely different shower and
hadronization frameworks \cite{Carloni:2010tw,Kulkarni:2024okx,Cohen:2015toa}. The same functional analysis could then be used to determine which generator-dependent structures of a dark shower survive into experimentally relevant jet observables.

A complementary direction is to repeat the analysis for different tunes within each generator. For example, variations between the Monash and CMS underlying-event tunes in \pythia{}, or between alternative \herwig{} underlying-event parameter sets, would allow one to distinguish differences associated with the generator framework itself from those induced by tuning the shower, MPI, colour-reconnection, and hadronization parameters \cite{Skands:2014pea,CMS:2019csb,CMS:2020dqt}. This KAN-based framework may also be useful in studies of generator decorrelation. As emphasized in \cite{Ghosh:2021hrh}, reducing sensitivity to one nominal generator variation does not by itself demonstrate that the underlying modeling uncertainty has been reduced. An explicit functional decomposition in this case will provide a more detailed language to understand this conundrum. One can then clearly identify which learned generator-sensitive directions are being removed by decorrelation, whether those directions are statistically supported, and whether they remain relevant when the event representation changes.

The main result of this work is therefore methodological as well as physical. The \pythia{}-\herwig{} difference in this dijet study does not correspond to one fixed observable direction propagated unchanged through the generator chain. Its functional anatomy changes with the stage of event generation, while only some early-stage information survives downstream. Additive KANs provide a particularly direct way of exposing this structure, separating persistent generator differences, genuinely non-persistent directions, and cases where finite statistical support prevents a reliable reweighting interpretation.

\section*{Acknowledgments}

The author thanks Phil Harris (MIT and IAIFI) and Stephen Mrenna (FNAL) for illuminating discussions at an early stage of this work and thanks the DPF 2026 conference for a stimulating environment. The numerical analysis in this work made extensive use of the open-source scientific Python ecosystem, and the author gratefully acknowledges its developers and maintainers. AI tools, including Anthropic’s Claude and OpenAI’s ChatGPT, were used to assist with code development and language refinement during the preparation of this work. This work is supported by U.S.A. National Science Foundation Award OAC-$2334265$.

\appendix

\section{Implementation details}
\label{app:technical}

\subsection{Production configuration}

The production sample uses \mg{} 2.9.24 for the common hard process, \pythia{} 8.312, and \herwig{} 7.3.0, with the NNPDF23\_lo\_as\_0130\_qed PDF set \cite{Ball:2013hta}. The physical
definitions of stages A-C are already discussed in \cref{sec:hardprocess}. For the non-perturbative and shower modelling, \pythia{} uses the Monash 2013 tune \cite{Skands:2014pea}, while \herwig{} $7.3$ is used with its default parameter configuration \cite{Bewick:2023tfi}. The stage definitions used in the main text correspond to the following generator-level logic:
\begin{center}
	\begin{tabular}{lccc}
		\toprule
		& Shower & Hadronization & MPI / underlying event \\
		\midrule
		A & on & off & off \\
		B & on & on & off \\
		C & on & on & on \\
		\bottomrule
	\end{tabular}
\end{center}

\subsection{Hard-event identity}

The staged analysis requires the six generator representations to correspond to the same underlying LHE event. HepMC event numbers \cite{Buckley:2019xhk} are therefore not used as the pairing key.

For \pythia{}, the production wrapper assigns
\begin{equation}
	\mathrm{hard\_event\_id}
	=
	\mathrm{block\ offset}
	+
	\mathrm{attempted\ LHE\ ordinal}
\end{equation}
before showering. The ordinal advances for every attempted \texttt{next()} call, including failed generator attempts, and the resulting identifier is stored explicitly in the HepMC event. Successful and failed identifiers are checked to partition the attempted LHE sequence exactly.

For \herwig{}, the hard-event identity is recovered from the complete MadGraph variation-weight information propagated through the event. A canonicalized fingerprint of this vector is required to identify exactly one entry in the LHE catalogue. No event-number, momentum-matching, or ordinal fallback is used. The Les Houches handler is run with the signed variable-weight option so that this information is preserved. Both identity constructions are audited before the common cohort is formed, and every generator-stage table is required to contain at most one entry for each hard-event identifier.

\subsection{Processed data and functional export}

Each of the six ($3$ stages each for the two event-generators) HepMC samples is processed independently with the jet reconstruction and selection of \cref{sec:jets}. The eight observables are stored together with the hard-event identifier, and the resulting tables are checked for finite features, unique identifiers, and the expected event range. The common cohort is constructed only after all six processed samples are available. Development/test membership and development-fold assignment are then generated deterministically from the hard-event identity, ensuring that all representations of one hard scattering remain together.

For each development/export fit, the additive KAN is stored as a finite numerical object containing the bias, input standardization, recorded source support, and the tabulated centered response functions $f_{s,i}(x_i)$. The exported representation is checked against the corresponding live KAN inside this support, together with the factorization in \cref{eq:weight_factorization}.

\section{Statistical definitions and validation}
\label{app:statistics}

\subsection{Hard-event clustering and the independent judge}

The cross-fitting procedure for the primary KAN models is defined in \cref{sec:training}. The same hard-event grouping is retained in all statistical resampling. Bootstrap replicas are formed by sampling \emph{hard-event identifiers} with replacement and including the associated
generator representations together. This preserves the pairing induced by the common matrix-element event.

Closure and transport are evaluated with a separate KAN judge. At a given target stage, this classifier is cross-fitted on the \emph{unweighted} \pythia{} and \herwig{} development samples. The resulting judge scores are kept fixed, and the transported or recomposed weights are applied only to the \pythia{} events when the weighted AUC is evaluated. Thus the unweighted and weighted comparisons use the same independent score function. The judge is not retrained after reweighting. For the final-test transport result, the judge applied to the final-test events is fitted using development data only, so the final-test sample enters no fitted quantity.

\subsection{Hard-event-clustered bootstrap}
\label{app:clustered_bootstrap}

Statistical uncertainties are estimated by bootstrap resampling at the level of the underlying hard event. This is necessary because the \pythia{} and \herwig{} representations carrying the same hard-event identifier originate from the same LHE matrix-element event and are therefore not statistically independent observations. The corresponding representations at stages A, B, and C are likewise tied to the same underlying hard scattering. A bootstrap replica is therefore constructed by sampling hard-event identifiers with replacement. Whenever a hard event is selected, all generator and stage representations required by the quantity being evaluated are included together. For example, in a stage-specific \pythia{}-\herwig{} comparison, the two generator representations of a selected hard event enter
or leave the bootstrap replica as a pair. For quantities involving several generator stages, the corresponding stage representations remain grouped under the same resampled hard-event identifier.

This prescription preserves the paired structure of the analysis. Resampling individual rows instead would treat the \pythia{} and \herwig{} evolutions of the same matrix-element event as independent measurements and could therefore underestimate the statistical uncertainty. The hard-event-clustered bootstrap instead asks how the result fluctuates when the underlying sample of hard scatterings is varied, while retaining the generator comparison associated with each scattering. The bootstrap intervals quoted for the Shapley allocations and the influence studies are obtained using this procedure.

\subsection{Effective sample size}

For non-negative source weights $w_i$, let
\begin{equation}
	\widetilde w_i
	=
	\frac{w_i}{\sum_j w_j}.
\end{equation}
we use the standard weight-based effective-sample-size estimator
\begin{equation}
	\ess
	=
	\frac{1}{\sum_i\widetilde w_i^2}
	=
	\frac{\left(\sum_i w_i\right)^2}
	{\sum_i w_i^2}.
	\label{eq:kish}
\end{equation}
Equal weights give $\ess=N$, whereas strong weight concentration can reduce the statistical content of a large event sample to that of only a few effectively independent events. This is particularly relevant for the transport analysis performed here. A source-stage factor may move the downstream \pythia{} distribution closer to \herwig{} according to the classifier judge, but such a change is physically meaningful only if it is supported by a sufficiently broad region of the source event population. Otherwise, the apparent closure can be driven by a small number of rare events receiving very large importance weights.

We therefore use $\ess/N$ as a direct measure of the statistical support of each transported factor. Values of order unity indicate that the reweighting is distributed over a substantial fraction of the source sample, whereas very small values signal that the result is dominated by a restricted tail of phase space. This provides information that the weighted AUC alone cannot, that two factors can produce similar changes in generator separation while having very different statistical reliability.

As a complementary diagnostic, we also inspect the cumulative normalized weight distribution. After ordering events by decreasing $\widetilde w_i$, the number of events required to accumulate $50\%$ of the total weight gives an intuitive measure of how localized the reweighting is. If hundreds or thousands of events are needed, the factor probes a broadly populated region. If only one or two events carry half of the weight, the corresponding transport cannot be interpreted as a robust population-level statement. Together, $\ess/N$ and the cumulative-weight diagnostic therefore separate genuine downstream persistence from apparent closure generated by poor source support.

\section{Functional Shapley implementation}
\label{app:shapley}

The subset and Shapley constructions are defined in \cref{sec:shapley_intro}. Numerically, all $2^8=256$ subsets are evaluated for each exported additive-KAN fit by exact recomposition of the stored functions as elaborated in algorithm \ref{alg:function_export}. For each seed and value function, the Shapley allocation is then evaluated from the complete subset lattice. Statistical intervals are obtained from $400$ hard-event-clustered bootstrap replicas per seed.

The allocation is first computed for the eight individual observables. The multiplicity, mass, and shape sectors shown in the main text are formed only afterwards by summing the appropriate feature-level Shapley values within each replica. The Shapley efficiency relation is checked for every seed and bootstrap realization. The largest residual in the present analysis is below
$4\times10^{-16}$. Negative allocations are allowed, they indicate that, averaged over subset orderings, adding the corresponding component reduces the chosen mathematical value function. For completeness, the numerical sector allocations used in \cref{fig:shapley_sectors} are listed in \cref{tab:shapley_full}. Note that because the Shapley decomposition is evaluated using the separate development or export fits of algorithm \ref{alg:function_export}, its full-subset separation need not coincide numerically with the cross-fitted OOF AUC quoted in table \ref{tab:staged_auc}.

\begin{table}[tbph]
	\centering
	\caption{Grouped Shapley allocations used in Analysis I. Brackets denote
		$95\%$ hard-event-clustered bootstrap intervals.}
	\label{tab:shapley_full}
	\small
	\setlength{\tabcolsep}{4pt}
\begin{tabular}{llccc}
	\toprule
	Stage & value function & multiplicity & mass & shape \\
	\midrule
	A & separation
	& $\ShSepAMult$ & $\ShSepAMass$ & $\ShSepAShape$ \\
	A & closure
	& $\ShCloAMult$ & $\ShCloAMass$ & $\ShCloAShape$ \\
	B & separation
	& $\ShSepBMult$ & $\ShSepBMass$ & $\ShSepBShape$ \\
	B & closure
	& $\ShCloBMult$ & $\ShCloBMass$ & $\ShCloBShape$ \\
	C & separation
	& $\ShSepCMult$ & $\ShSepCMass$ & $\ShSepCShape$ \\
	C & closure
	& $\ShCloCMult$ & $\ShCloCMass$ & $\ShCloCShape$ \\
	\bottomrule
\end{tabular}
\end{table}
\section{Weight support and influence diagnostics}
\label{app:weights}

The effective-sample and cumulative-weight diagnostics in \cref{sec:mass_transport} establish that the stage A mass factors do not support a reliable population-level transport measurement. Here we elaborate on the additional influence tests to determine whether this behaviour is driven by individual rare events or instead reflects a more persistent lack of statistical support.

\subsection{Tail support}

The concentration discussed in \cref{sec:mass_transport} can be traced to the sparsely populated stage-A mass tails. For a representative choice of random seed, the most influential leading-mass event lies at approximately $\LeadMassDominantMassGeV~\mathrm{GeV}$ and carries about $\LeadMassDominantWeightPct$ of the total component weight. The corresponding subleading-mass event carries about $\SubMassDominantWeightPct$. Thus the poor support is associated with a very small number of rare tail events rather than with a broad mismatch across the full source distribution.

This behaviour is qualitatively different from the multiplicity factors. For both leading and subleading multiplicity, the five largest weights together carry less than one percent of the corresponding total weight, and hundreds of development events are required to accumulate half of the reweighted measure. Agreement across training seeds is therefore not by itself evidence of statistical support, rather several independently trained models can consistently identify the same sparsely populated tail.

\subsection{Influence and targeted refitting tests}

We first condition the hard-event-clustered bootstrap on whether the highest-weight source event is present. For leading mass, the diagnostic weighted AUC is approximately $\LeadMassAUCWithDominant$ when the dominant event is present and $\LeadMassAUCWithoutDominant$ when it is absent. For subleading mass, the corresponding values are $\SubMassAUCWithDominant$ and $\SubMassAUCWithoutDominant$. The leading-multiplicity control instead gives $\LeadMultAUCWithDominant$ and $\LeadMultAUCWithoutDominant$, respectively. The mass-weighted diagnostics are therefore sensitive to the presence of individual high-weight events, whereas the supported multiplicity transport is substantially more stable under the same test.

As a complementary test, we retrain the additive KAN after deliberately removing the most influential stage A events. The two mass factors respond differently. For subleading mass, removing the dominant event changes the range of the centered response from approximately
$\SubMassRangeOriginal$ to $\SubMassRangeDropOne$, showing that the learned tail is directly sensitive to that particular event. For leading mass, removing the five most influential events changes the response range only from approximately $\LeadMassRangeOriginal$ to
$\LeadMassRangeDropFive$. In this case the large weights migrate to the next available events in the sparsely populated tail, and even after five removals the effective-sample fraction remains only $\ess/N\simeq\LeadMassESSDropFive$.

The two mass factors therefore fail the support test in different ways. The subleading-mass response contains a strongly event-sensitive tail, while the leading-mass response exhibits a more persistent concentration across the available high-mass tail. Neither case provides a statistically supported standalone transport weight. More generally, these tests illustrate why an explicit and reproducible response function need not correspond to a statistically usable importance weight. The functional structure may be well defined even when the available source sample does not provide sufficient support for its population-level transport.

\bibliographystyle{jhep}
\bibliography{references}

\end{document}

%% file: numbers.tex
\newcommand{\NCandidate}{175000}
\newcommand{\NCommon}{11086}
\newcommand{\CommonEff}{6.33\%}
\newcommand{\NDev}{8833}
\newcommand{\NTest}{2253}

\newcommand{\AAddA}{0.9787}
\newcommand{\ADeepA}{0.9924}
\newcommand{\ADeltaA}{0.0137}
\newcommand{\AAddB}{0.5933}
\newcommand{\ADeepB}{0.5941}
\newcommand{\ADeltaB}{0.0008}
\newcommand{\AAddC}{0.6061}
\newcommand{\ADeepC}{0.6062}
\newcommand{\ADeltaC}{0.0001}

\newcommand{\DevBUnw}{0.5943}
\newcommand{\DevBLeadMult}{0.5132}
\newcommand{\TestBUnw}{0.5936}
\newcommand{\TestBLeadMult}{0.5249}
\newcommand{\DevCUnw}{0.6065}
\newcommand{\DevCLeadMult}{0.5459}
\newcommand{\TestCUnw}{0.6167}
\newcommand{\TestCLeadMult}{0.5619}
\newcommand{\LeadMultESSDev}{0.1080}
\newcommand{\LeadMultESSTest}{0.1022}

\newcommand{\SubMultB}{0.5521}
\newcommand{\SubMultC}{0.5566}
\newcommand{\SubMultESS}{0.1408}
\newcommand{\LeadPTDB}{0.6064}
\newcommand{\LeadPTDC}{0.6170}

\newcommand{\SubPTDB}{0.6006}
\newcommand{\SubPTDC}{0.6129}

\newcommand{\LeadGirthB}{0.5934}
\newcommand{\LeadGirthC}{0.6055}

\newcommand{\SubGirthB}{0.5928}
\newcommand{\SubGirthC}{0.6059}

\newcommand{\LeadMassESS}{4.38\times10^{-4}}
\newcommand{\SubMassESS}{1.50\times10^{-4}}

\newcommand{\ShSepAMult}{0.4050\,[0.4021,0.4078]}
\newcommand{\ShSepAMass}{-0.0005\,[-0.0034,0.0022]}
\newcommand{\ShSepAShape}{0.0751\,[0.0720,0.0789]}
\newcommand{\ShCloAMult}{0.1796\,[0.1403,0.2186]}
\newcommand{\ShCloAMass}{0.0286\,[0.0072,0.0538]}
\newcommand{\ShCloAShape}{-0.0001\,[-0.0048,0.0044]}
\newcommand{\ShSepBMult}{0.0039\,[0.0024,0.0056]}
\newcommand{\ShSepBMass}{0.0498\,[0.0442,0.0554]}
\newcommand{\ShSepBShape}{0.0478\,[0.0415,0.0544]}
\newcommand{\ShCloBMult}{-0.0024\,[-0.0031,-0.0013]}
\newcommand{\ShCloBMass}{0.0537\,[0.0514,0.0553]}
\newcommand{\ShCloBShape}{0.0378\,[0.0341,0.0392]}
\newcommand{\ShSepCMult}{0.0365\,[0.0308,0.0420]}
\newcommand{\ShSepCMass}{0.0205\,[0.0148,0.0257]}
\newcommand{\ShSepCShape}{0.0554\,[0.0488,0.0625]}
\newcommand{\ShCloCMult}{0.0352\,[0.0304,0.0390]}
\newcommand{\ShCloCMass}{0.0133\,[0.0104,0.0153]}
\newcommand{\ShCloCShape}{0.0519\,[0.0480,0.0561]}

\newcommand{\LeadMassHalfWeightEventsSeedZero}{2}

\newcommand{\SubMassHalfWeightEventsSeedZero}{1}

\newcommand{\LeadMassDominantWeightPct}{49.1\%}
\newcommand{\LeadMassDominantMassGeV}{39.0}
\newcommand{\SubMassDominantWeightPct}{87.1\%}
\newcommand{\LeadMassAUCWithDominant}{0.8841}
\newcommand{\LeadMassAUCWithoutDominant}{0.8430}
\newcommand{\SubMassAUCWithDominant}{0.8970}
\newcommand{\SubMassAUCWithoutDominant}{0.7642}
\newcommand{\LeadMultAUCWithDominant}{0.5142}
\newcommand{\LeadMultAUCWithoutDominant}{0.5139}
\newcommand{\SubMassRangeOriginal}{19.33}
\newcommand{\SubMassRangeDropOne}{17.52}
\newcommand{\LeadMassRangeOriginal}{22.36}
\newcommand{\LeadMassRangeDropFive}{22.52}
\newcommand{\LeadMassESSDropFive}{0.0042}

%% file: references.bib
@article{Bierlich:2022pfr,
	author = "Bierlich, Christian and others",
	title = "{A comprehensive guide to the physics and usage of PYTHIA 8.3}",
	eprint = "2203.11601",
	archivePrefix = "arXiv",
	primaryClass = "hep-ph",
	reportNumber = "LU-TP 22-16, MCNET-22-04, FERMILAB-PUB-22-227-SCD",
	doi = "10.21468/SciPostPhysCodeb.8",
	journal = "SciPost Phys. Codeb.",
	volume = "2022",
	pages = "8",
	year = "2022"
}

@article{Bewick:2023tfi,
	author = "Bewick, Gavin and others",
	title = "{Herwig 7.3 release note}",
	eprint = "2312.05175",
	archivePrefix = "arXiv",
	primaryClass = "hep-ph",
	reportNumber = "CERN-TH-2023-223, HERWIG-2023-01, KA-TP-28-2023, MCnet-23-19, IPPP/23/66",
	doi = "10.1140/epjc/s10052-024-13211-9",
	journal = "Eur. Phys. J. C",
	volume = "84",
	number = "10",
	pages = "1053",
	year = "2024"
}

@article{Andersson:1983ia,
	author = "Andersson, Bo and Gustafson, G. and Ingelman, G. and Sjostrand, T.",
	title = "{Parton Fragmentation and String Dynamics}",
	reportNumber = "LU-TP-83-10",
	doi = "10.1016/0370-1573(83)90080-7",
	journal = "Phys. Rept.",
	volume = "97",
	pages = "31--145",
	year = "1983"
}

@article{Webber:1983if,
	author = "Webber, B. R.",
	title = "{A QCD Model for Jet Fragmentation Including Soft Gluon Interference}",
	reportNumber = "CERN-TH-3713",
	doi = "10.1016/0550-3213(84)90333-X",
	journal = "Nucl. Phys. B",
	volume = "238",
	pages = "492--528",
	year = "1984"
}

@article{Alwall:2014hca,
	author = "Alwall, J. and Frederix, R. and Frixione, S. and Hirschi, V. and Maltoni, F. and Mattelaer, O. and Shao, H. -S. and Stelzer, T. and Torrielli, P. and Zaro, M.",
	title = "{The automated computation of tree-level and next-to-leading order differential cross sections, and their matching to parton shower simulations}",
	eprint = "1405.0301",
	archivePrefix = "arXiv",
	primaryClass = "hep-ph",
	reportNumber = "CERN-PH-TH-2014-064, CP3-14-18, LPN14-066, MCNET-14-09, ZU-TH-14-14",
	doi = "10.1007/JHEP07(2014)079",
	journal = "JHEP",
	volume = "07",
	pages = "079",
	year = "2014"
}

@article{Cacciari:2008gp,
	author = "Cacciari, Matteo and Salam, Gavin P. and Soyez, Gregory",
	title = "{The anti-$k_t$ jet clustering algorithm}",
	eprint = "0802.1189",
	archivePrefix = "arXiv",
	primaryClass = "hep-ph",
	reportNumber = "LPTHE-07-03",
	doi = "10.1088/1126-6708/2008/04/063",
	journal = "JHEP",
	volume = "04",
	pages = "063",
	year = "2008"
}

@article{Cacciari:2011ma,
	author = "Cacciari, Matteo and Salam, Gavin P. and Soyez, Gregory",
	title = "{FastJet User Manual}",
	eprint = "1111.6097",
	archivePrefix = "arXiv",
	primaryClass = "hep-ph",
	reportNumber = "CERN-PH-TH-2011-297",
	doi = "10.1140/epjc/s10052-012-1896-2",
	journal = "Eur. Phys. J. C",
	volume = "72",
	pages = "1896",
	year = "2012"
}

@article{Gallicchio:2011xq,
	author = "Gallicchio, Jason and Schwartz, Matthew D.",
	title = "{Quark and Gluon Tagging at the LHC}",
	eprint = "1106.3076",
	archivePrefix = "arXiv",
	primaryClass = "hep-ph",
	doi = "10.1103/PhysRevLett.107.172001",
	journal = "Phys. Rev. Lett.",
	volume = "107",
	pages = "172001",
	year = "2011"
}

@article{Larkoski:2014pca,
	author = "Larkoski, Andrew J. and Thaler, Jesse and Waalewijn, Wouter J.",
	title = "{Gaining (Mutual) Information about Quark/Gluon Discrimination}",
	eprint = "1408.3122",
	archivePrefix = "arXiv",
	primaryClass = "hep-ph",
	reportNumber = "MIT--CTP-4572, NIKHEF-2014-026",
	doi = "10.1007/JHEP11(2014)129",
	journal = "JHEP",
	volume = "11",
	pages = "129",
	year = "2014"
}

@article{Cranmer:2015bka,
	author = "Cranmer, Kyle and Pavez, Juan and Louppe, Gilles",
	title = "{Approximating Likelihood Ratios with Calibrated Discriminative  Classifiers}",
	eprint = "1506.02169",
	archivePrefix = "arXiv",
	primaryClass = "stat.AP",
	month = "6",
	year = "2015"
}

@article{Andreassen:2019nnm,
	author = "Andreassen, Anders and Nachman, Benjamin",
	title = "{Neural Networks for Full Phase-space Reweighting and Parameter Tuning}",
	eprint = "1907.08209",
	archivePrefix = "arXiv",
	primaryClass = "hep-ph",
	doi = "10.1103/PhysRevD.101.091901",
	journal = "Phys. Rev. D",
	volume = "101",
	number = "9",
	pages = "091901",
	year = "2020"
}

@article{Furuichi:2025generator,
	author = "Furuichi, Amon and Lim, Sung Hak and Nojiri, Mihoko M.",
	title = "{Reweighting and analysing event generator systematics by neural networks on high-level features}",
	eprint = "2503.01452",
	archivePrefix = "arXiv",
	primaryClass = "hep-ph",
	reportNumber = "CTPU-PTC-25-07",
	doi = "10.1007/JHEP07(2025)111",
	journal = "JHEP",
	volume = "07",
	pages = "111",
	year = "2025"
}

@article{Ghosh:2021hrh,
	author = "Ghosh, Aishik and Nachman, Benjamin",
	title = "{A cautionary tale of decorrelating theory uncertainties}",
	eprint = "2109.08159",
	archivePrefix = "arXiv",
	primaryClass = "hep-ph",
	doi = "10.1140/epjc/s10052-022-10012-w",
	journal = "Eur. Phys. J. C",
	volume = "82",
	number = "1",
	pages = "46",
	year = "2022"
}

@article{Kolmogorov:1957,
  author = {Kolmogorov, A. N.},
  title = {On the representation of continuous functions of many variables by superposition of continuous functions of one variable and addition},
  journal = {Dokl. Akad. Nauk SSSR},
  volume = {114},
  number = {5},
  pages = {953--956},
  year = {1957}
}

@article{Arnold:1957,
  author = {Arnol'd, V. I.},
  title = {On functions of three variables},
  journal = {Dokl. Akad. Nauk SSSR},
  volume = {114},
  number = {4},
  pages = {679--681},
  year = {1957}
}

@article{Liu:2024kan,
	author = "Liu, Ziming and Wang, Yixuan and Vaidya, Sachin and Ruehle, Fabian and Halverson, James and Solja{\v{c}}i{\'c}, Marin and Hou, Thomas Y. and Tegmark, Max",
	title = "{KAN: Kolmogorov-Arnold Networks}",
	eprint = "2404.19756",
	archivePrefix = "arXiv",
	primaryClass = "cs.LG",
	month = "4",
	year = "2024"
}

@article{Erdmann:2025kan,
	author = {Erdmann, Johannes and Mausolf, Florian and Sp{\"a}h, Jan Lukas},
	title = "{KAN We Improve on HEP Classification Tasks? Kolmogorov{\textendash}Arnold Networks Applied to an LHC Physics Example}",
	eprint = "2408.02743",
	archivePrefix = "arXiv",
	primaryClass = "hep-ph",
	doi = "10.1007/s41781-025-00138-3",
	journal = "Comput. Softw. Big Sci.",
	volume = "9",
	number = "1",
	pages = "9",
	year = "2025"
}

@article{Abasov:2024kan,
	author = "Abasov, E. E. and Volkov, P. V. and Vorotnikov, G. A. and Dudko, L. V. and Zaborenko, A. D. and Iudin, E. S. and Markina, A. A. and Perfilov, M. A.",
	title = "{Application of Kolmogorov{\textendash}Arnold Networks in High Energy Physics}",
	eprint = "2409.01724",
	archivePrefix = "arXiv",
	primaryClass = "hep-ph",
	doi = "10.3103/S0027134924702035",
	journal = "Moscow Univ. Phys. Bull.",
	volume = "79",
	number = "Suppl 2",
	pages = "S585--S590",
	year = "2024"
}

@misc{pykanRepo,
  author = {Liu, Ziming and collaborators},
  title = {pyKAN: Kolmogorov-Arnold Networks},
  howpublished = {\url{https://github.com/KindXiaoming/pykan}},
  note = {Official implementation of Kolmogorov--Arnold networks},
  year = {2024}
}

@incollection{Shapley:1953,
  author = {Shapley, Lloyd S.},
  title = {A Value for $n$-Person Games},
  booktitle = {Contributions to the Theory of Games II},
  editor = {Kuhn, H. W. and Tucker, A. W.},
  publisher = {Princeton University Press},
  pages = {307--317},
  year = {1953}
}

@misc{Anatomy_pythia_herwig,
	author       = {Chattopadhyay, Arghya},
	title        = {Pythia\_Herwig\_anatomy: code and reproducibility package for ``Functional anatomy of Pythia--Herwig differences with Kolmogorov--Arnold networks''},
	year         = {2026},
	howpublished = {\url{https://github.com/chattopadhyayA/Pythia_Herwig_anatomy}},
	note         = {GitHub repository}
}

@article{Bewick:2021nhc,
	author = "Bewick, Gavin and Ferrario Ravasio, Silvia and Richardson, Peter and Seymour, Michael H.",
	title = "{Initial state radiation in the Herwig 7 angular-ordered parton shower}",
	eprint = "2107.04051",
	archivePrefix = "arXiv",
	primaryClass = "hep-ph",
	reportNumber = "CERN-TH-2021-103, OUTP-21-18P, MCnet-21-13, IPPP/21/05",
	doi = "10.1007/JHEP01(2022)026",
	journal = "JHEP",
	volume = "01",
	pages = "026",
	year = "2022"
}

@article{Skands:2014pea,
	author = "Skands, Peter and Carrazza, Stefano and Rojo, Juan",
	title = "{Tuning PYTHIA 8.1: the Monash 2013 Tune}",
	eprint = "1404.5630",
	archivePrefix = "arXiv",
	primaryClass = "hep-ph",
	reportNumber = "CERN-PH-TH-2014-069, MCNET-14-08, OUTP-14-05P",
	doi = "10.1140/epjc/s10052-014-3024-y",
	journal = "Eur. Phys. J. C",
	volume = "74",
	number = "8",
	pages = "3024",
	year = "2014"
}

@article{Alwall:2006yp,
	author = "Alwall, J. and others",
	title = "{A Standard format for Les Houches event files}",
	eprint = "hep-ph/0609017",
	archivePrefix = "arXiv",
	reportNumber = "FERMILAB-PUB-06-337-T, CERN-LCGAPP-2006-03",
	doi = "10.1016/j.cpc.2006.11.010",
	journal = "Comput. Phys. Commun.",
	volume = "176",
	pages = "300--304",
	year = "2007"
}

@article{kingma2014adam,
	title={Adam: A method for stochastic optimization},
	author={Kingma, Diederik P and Ba, Jimmy},
	journal={arXiv preprint arXiv:1412.6980},
	year={2014}
}

@article{Ball:2013hta,
	author = "Ball, Richard D. and Bertone, Valerio and Carrazza, Stefano and Del Debbio, Luigi and Forte, Stefano and Guffanti, Alberto and Hartland, Nathan P. and Rojo, Juan",
	collaboration = "NNPDF",
	title = "{Parton distributions with QED corrections}",
	eprint = "1308.0598",
	archivePrefix = "arXiv",
	primaryClass = "hep-ph",
	reportNumber = "EDINBURGH-2013-20, FR-PHENO-2013-008, CERN-PH-TH-2013-075, Edinburgh 2013/20, IFUM-1014-FT, FR-PHENO-2013-008,
	CERN-PH-TH/2013-075",
	doi = "10.1016/j.nuclphysb.2013.10.010",
	journal = "Nucl. Phys. B",
	volume = "877",
	pages = "290--320",
	year = "2013"
}

@article{Buckley:2019xhk,
	author = {Buckley, Andy and Ilten, Philip and Konstantinov, Dmitri and L{\"o}nnblad, Leif and Monk, James and Pokorski, Witold and Przedzinski, Tomasz and Verbytskyi, Andrii},
	title = "{The HepMC3 event record library for Monte Carlo event generators}",
	eprint = "1912.08005",
	archivePrefix = "arXiv",
	primaryClass = "hep-ph",
	reportNumber = "MPP-2019-258, MCNET-19-27, LU-TP 19-58",
	doi = "10.1016/j.cpc.2020.107310",
	journal = "Comput. Phys. Commun.",
	volume = "260",
	pages = "107310",
	year = "2021"
}

@article{Kulkarni:2024okx,
	author = {Kulkarni, Suchita and Masouminia, M. R. and Pl{\"a}tzer, Simon and Stafford, Dominic},
	title = "{Dark sector showers and hadronisation in Herwig 7}",
	eprint = "2408.10044",
	archivePrefix = "arXiv",
	primaryClass = "hep-ph",
	reportNumber = "IPPP/24/54, IPPP/24/54; PUBDB-2024-06237",
	doi = "10.1140/epjc/s10052-024-13587-8",
	journal = "Eur. Phys. J. C",
	volume = "84",
	number = "11",
	pages = "1210",
	year = "2024"
}

@article{Carloni:2010tw,
	author = "Carloni, Lisa and Sjostrand, Torbjorn",
	title = "{Visible Effects of Invisible Hidden Valley Radiation}",
	eprint = "1006.2911",
	archivePrefix = "arXiv",
	primaryClass = "hep-ph",
	reportNumber = "LU-TP-10-17, MCNET-10-11",
	doi = "10.1007/JHEP09(2010)105",
	journal = "JHEP",
	volume = "09",
	pages = "105",
	year = "2010"
}

@article{Cohen:2015toa,
	author = "Cohen, Timothy and Lisanti, Mariangela and Lou, Hou Keong",
	title = "{Semivisible Jets: Dark Matter Undercover at the LHC}",
	eprint = "1503.00009",
	archivePrefix = "arXiv",
	primaryClass = "hep-ph",
	doi = "10.1103/PhysRevLett.115.171804",
	journal = "Phys. Rev. Lett.",
	volume = "115",
	number = "17",
	pages = "171804",
	year = "2015"
}

@article{CMS:2019csb,
	author = "Sirunyan, Albert M and others",
	collaboration = "CMS",
	title = "{Extraction and validation of a new set of CMS PYTHIA8 tunes from underlying-event measurements}",
	eprint = "1903.12179",
	archivePrefix = "arXiv",
	primaryClass = "hep-ex",
	reportNumber = "CMS-GEN-17-001, CERN-EP-2019-007",
	doi = "10.1140/epjc/s10052-019-7499-4",
	journal = "Eur. Phys. J. C",
	volume = "80",
	number = "1",
	pages = "4",
	year = "2020"
}

@article{CMS:2020dqt,
	author = "Sirunyan, Albert M and others",
	collaboration = "CMS",
	title = "{Development and validation of HERWIG 7 tunes from CMS underlying-event measurements}",
	eprint = "2011.03422",
	archivePrefix = "arXiv",
	primaryClass = "hep-ex",
	reportNumber = "CMS-GEN-19-001, CERN-EP-2020-182",
	doi = "10.1140/epjc/s10052-021-08949-5",
	journal = "Eur. Phys. J. C",
	volume = "81",
	number = "4",
	pages = "312",
	year = "2021"
}
